\pdfoutput=1

\documentclass[12pt]{article}
\usepackage{lipsum}

\usepackage{latexsym} 
\usepackage{verbatim}
\usepackage{tikz}
\usetikzlibrary{matrix}
\usepackage{color}
\usepackage{graphicx,amssymb,amsfonts,amsmath,amssymb,amscd,amstext, mathrsfs}
\usepackage{graphicx} 
\usepackage{bbm}
\usepackage{dsfont}
\usepackage{xcolor}
\usepackage[colorlinks=true, 
citecolor=black, linkcolor=black, allcolors=black]{hyperref}   
\usepackage{amsmath}
\usepackage{empheq}
\usepackage{hyperref}

\newlength\dlf

\def\be{\begin{eqnarray}}
\def\ee{\end{eqnarray}}
\def \bea {\begin{equation}}
\def \eea {\end{equation}}

\def \nn {\nonumber}

\def \rr {\raise.35ex\hbox{\small $\prime$}\kern-.17em{\mbox{\large $\imath$}}}

\def \dels {\partial\kern-.5em / \kern.5em}
\def \As {{A\kern-.5em / \kern.5em}}
\def \Ds {D\kern-.7em / \kern.5em}

\def \k {\kappa}

\def \z {(0)}

\def\frac#1#2{{#1\over #2}}

\newcommand{\<}{\langle}
\renewcommand{\>}{\rangle}

\def \iffa {\iffalse} 

\def \ed  {\end{document}}

\def\nn{\nonumber}
\def \k {\kappa}

\def \ed  {\end{document}}

\DeclareFontShape{OT1}{cmr}{mx}{n}{<->cmr10}{}

\def \z{\bar z}

\usepackage{cite}

\begin{document}

\begin{titlepage}

\begin{flushright} 
\end{flushright}

\begin{center} 
\vspace{2.5cm}  

{\fontsize{15pt}{0pt}{\bf Toward Null State Equations in $d>2$ 
}}
\vspace{1.7cm}  
\\
 {\fontsize{12pt}{0pt}{Kuo-Wei Huang 
}}   
\\ 
\vspace{0.3cm} 
{\fontsize{11.5pt}{0pt}{\it Mathematical Sciences, University of Southampton, \\
Highfield, Southampton SO17 1BJ, UK}}\\
\end{center}
\vspace{0.8cm} 

\begin{center} 
{\bf Abstract}
\end{center} 
\vspace{-0.3cm} 
{\noindent 
 In two-dimensional CFTs with a large central charge, the level-two BPZ equation governs the heavy-light scalar four-point correlator when the light probe scalar has dimension $h= - {1\over 2}$; the corresponding linear ordinary differential equation can be recast into a schematic form $x^2 u_{xx}+u=0$. In this paper, we make an observation that in a class of four-dimensional CFTs with a large central charge, the heavy-light scalar correlator in the near-lightcone limit  obeys a similar equation, $x^3 u_{xxxy}+u=0$, when the light scalar has dimension $\Delta=-1$.  We focus on the multi-stress tensor sector of the theory and also discuss the corresponding equations for the cases with $\Delta = -2, -3$. The solutions to these linear partial differential equations in higher dimensions are shown, after a suitable change of variables, to reproduce the near-lightcone correlators previously obtained via holography and the conformal bootstrap.}

\end{titlepage}

\addtolength{\parskip}{1 ex}
\jot=2 ex


\subsection*{1. Introduction and Summary}

It is safe to say that our understanding of generic Quantum Field Theory (QFT) is still in its infancy.  
In particular, computing physical observables at strong coupling presents a primary difficulty.  
As a special class of QFTs  that describe the critical phenomena at second-order phase transitions, Conformal Field Theory (CFT) has the convergent expansions in conformal blocks, allowing rigorous computations. The extra symmetries in CFT lead to  non-trivial constraints, sometimes powerful enough to solve the theory.  Almost forty years ago,  Belavin, Polyakov, and Zamolodchikov (BPZ) discovered that the $d=2$ conformal correlators in Virasoro minimal models obey a set of null-state linear differential equations \cite{BELAVIN1984333}, beyond the usual Ward identities.  As a consequence of the enhanced infinite-dimensional Virasoro symmetry, these linear differential equations allow us to compute the correlators without using perturbative Lagrangian techniques.  Although the landscape of $d=2$ CFTs remains largely uncharted, the present paper aims to search for BPZ-type equations in higher dimensional CFTs at large central charge (large N), focusing on a scalar four-point correlator in the {\it near-lightcone} regime.
  
Certainly, in $d>2$ the conformal group is finite dimensional and the factorization between a holomorphic and an anti-holomorphic part is a special feature in $d=2$. The Virasoro-algebra related techniques cannot be extended to higher dimensions in general. 
There is no {\it a priori} reason either to expect that large-N CFT correlators in a kinematical (near-lightcone) limit would satisfy any BPZ-type equations in $d>2$, unless an effective symmetry emerges in a system with large degrees of freedom or an effective action near the lightcone can be used. Very often when a question is difficult to address  due to a lack of clues, trying to identify the final answer directly may help  reconstruct an underlying reason afterward -- this is the approach we adopt in the present work.   

We will discuss a set of linear differential equations in $d=4$, inspired by crossing symmetry and similarity with the level-2  BPZ equation, and verify their agreement with results obtained using other techniques.  Furthermore, we will demonstrate the effectiveness of these new equations by presenting results that were previously unattainable. Our approach is not entirely satisfactory due to the lack of a first-principle derivation for these equations.\footnote{By a first-principle approach, we are referring to a CFT derivation that begins with $TT$ and $T\cal {O}$ operator product expansions (OPEs).}  There is, therefore, no fundamentally new organizing principle that will be established in this paper.  However, it is always interesting to try to infer the final answer via  pattern recognition, and to observe that simple equations govern intricate recursive structure. As a part of our motivation, we hope that these $d=4$ differential equations we observe may serve as an impetus to develop a rigorous framework in the future. 

Studying four-point conformal correlators in the near-lightcone limit is a very useful starting point toward a better understanding of  higher dimensional CFTs more generally.  In such an effective limit, a vast simplification occurs, yet preserving the richness and intricacy of CFT dynamics. One can analytically solve the corresponding crossing equations, $i.e.$, the lightcone bootstrap equations, and  crossing symmetry requires the existence of large spin double-twist operators \cite{Fitzpatrick:2012yx,Komargodski:2012ek}. See \cite{Alday:2007mf} for an earlier discussion and a recent work \cite{Pal:2022vqc} for a rigorous treatment. Recall that a local operator can be characterized by its twist $\tau=\Delta- J$, where $\Delta, J$ are the dimension and spin.   The near-lightcone correlator encodes information about the leading-twist sector of the theory.
In this work, we are interested in the primaries  formed by the product of $n$ stress tensors, denoted schematically as $T^n$, with the lowest twist $\tau_{\rm min}= n (d-2)$. These leading-twist multi-stress tensors are built from stress tensors and possible derivatives with no contracted Lorentz indices, ensuring the twist to be minimal. In two dimensions, the Virasoro algebra determines the resummation of leading-twist $T^n$ operators which is given by the Virasoro vacuum block, whose structure and various applications have been extensively studied in the literature, $e.g.$, \cite{zamolodchikov1984conformal, zamolodchikov2, Harlow:2011ny, Hartman:2013mia, Litvinov:2013sxa, Fitzpatrick:2014vua, Asplund:2014coa, Roberts:2014ifa,  Hijano:2015rla, Fitzpatrick:2015zha, Perlmutter:2015iya,  Hijano:2015qja, Fitzpatrick:2015foa,  Beccaria:2015shq, Fitzpatrick:2015dlt, Alkalaev:2015fbw,  Fitzpatrick:2016thx, Banerjee:2016qca,  Anous:2016kss,   Fitzpatrick:2016ive, Chen:2016cms, Chen:2016dfb,  Maloney:2016kee,  Fitzpatrick:2016mtp,  Chen:2017yze, Cho:2017oxl, Cho:2017fzo,  Galliani:2017jlg, Bombini:2017sge,  Kusuki:2018nms,  Bombini:2018jrg, Cotler:2018zff,  Kusuki:2018wpa,  Collier:2018exn, Kulaxizi:2018dxo, Chang:2018nzm,  Giusto:2018ovt, Anous:2019yku, Kusuki:2019gjs,  Besken:2019bsu,  Haehl:2019eae,  Besken:2019jyw, Collier:2019weq, Alkalaev:2020kxz, Anous:2020vtw,  Das:2020fhs,  Nguyen:2021jja, Nguyen:2022xsw, Benjamin:2023uib}. 

In this paper, we restrict our attention to a class of $d=4$ CFTs at large central charge $C_T$ with a large gap in the spectrum of higher-spin single-trace operators. 
These conditions are required for a CFT to have a local gravitational dual description in Anti-de Sitter space (AdS), as conjectured in \cite{Heemskerk:2009pn}. See, $e.g.$,   \cite{Fitzpatrick:2010zm, Camanho:2014apa, Hartman:2015lfa, Komargodski:2016gci, Afkhami-Jeddi:2016ntf, Kulaxizi:2017ixa, Li:2017lmh, Meltzer:2017rtf, Afkhami-Jeddi:2018apj,  Belin:2019mnx, Kologlu:2019bco, Caron-Huot:2021enk} for related discussions. At large $C_T$, a natural observable is the ``heavy-light" four-point correlator, $\langle {\cal O}_H {\cal O}_L {\cal O}_L  {\cal O}_H\rangle$. The heavy scalars have dimensions $\Delta_H$ of order $C_T$ which, importantly, compensate for the large $C_T$ suppression. 
In such a ``semiclassical" limit, the $\Delta_H$ dependence enters the correlator only through the fixed ratio $\mu \sim {\Delta_H\over C_T}$ as $C_T\to \infty$. The light probe scalars, on the other hand, have dimensions $\Delta_L\equiv \Delta$ of order one.     
Focusing on the contributions from multi-stress tensors, the $d=4$ correlator in the near-lightcone limit $\bar z \to 0$ 
can be represented as
\begin{align}
\label{defineF}
F(z, \bar z)\equiv\lim_{\bar z \to 0} {\langle {\cal O}_H (\infty) {\cal O}_L(1) {\cal O}_L(1-z, 1-\z)  {\cal O}_H(0)\rangle \over   \langle {\cal O}_H(\infty)  {\cal O}_H (0) \rangle \langle {\cal O}_L(1) {\cal O}_L (1-z, 1-\z) \rangle}\Big|_{T^n} 
= \sum^{\infty}_{n=0} G_n (\Delta, z) (\mu \bar z)^n .  
\end{align} 
The functions $G_n (\Delta, z)$ are generally challenging to compute for $n \geq 2$.  The power of the parameter $\mu$ counts the number of exchanged stress-tensor operators. When $n=0$, we have $G_0 =1$ for the vacuum correlator.  The $n=1 $ case corresponds to the single-stress tensor exchange which is universally fixed by the Ward identity.  Note that the parameter $\mu$ can be absorbed into $\bar z$ in the near-lightcone expansion \eqref{defineF}.  We will scale $\z$ so that the $d=4$ expressions do not show an explicit $\mu$.  In $d=2$, $\mu$ is the expansion parameter since the holomorphic function $F(z)$ has no $\bar z$ dependence.

Using the AdS/CFT correspondence \cite{Maldacena:1997re, Gubser:1998bc, Witten:1998qj}, one can compute the heavy-light four-point correlator at large $C_T$ as a thermal two-point scalar correlator in a black-hole background created by the heavy operators.   The Hawking temperature of a black hole can be related to the CFT parameter $\mu$. The multi-stress tensor exchanges  are dual to multi-graviton exchanges in AdS.  In the context of AdS$_3$/CFT$_2$, the bulk-to-boundary propagator with the Banados-Teitelboim-Zanelli (BTZ) black hole \cite{Banados:1992wn} has a closed-form expression \cite{Keski-Vakkuri:1998gmz}, allowing for various analytic computations. Moreover, due to the Virasoro symmetry in the dual $d=2$ CFTs, it is anticipated that higher-curvature corrections to the AdS$_3$ gravitational action do not affect the boundary  correlators. In higher dimensions, the correlators depend on the specific theory. It is, however, interesting to investigate whether there is a subsector of the theory that exhibits a  more robust structure.    In \cite{Fitzpatrick:2019zqz}, it was observed that the leading-twist sector of multi-stress tensors in the dual $d>2$ CFTs is insensitive to higher-curvature corrections. This observation was made by considering a minimally coupled probe scalar in a black-hole background in AdS$_{D>3}$.
The near-lightcone correlators at large $C_T$ take the same form as that computed in pure Einstein gravity. 
This statement goes beyond the geodesic approximation.\footnote{In addition to the $T^n$ operators, there are contributions from the double-trace operators made from two light scalars and derivatives.   Their contributions depend on an interior boundary condition \cite{Fitzpatrick:2019zqz}.  In this work we focus on the contributions from multi-stress tensors.}  
Furthermore, it was observed that certain simplifications of the correlator occur at special negative integers of $\Delta$\cite{Fitzpatrick:2019zqz}
\begin{align}
\label{specialDelta}
\Delta =-1,-2, \cdots, - {d+2\over 2} \ , ~~~~~ d = 4,6,8, \cdots
\end{align} 
where closed-form expressions summing over {\it all} multi-stress tensor exchanges can be found (based on a computation with a planar black hole).  The radius of convergence becomes infinite at these special values of $\Delta$.  
The motivation of the present work stems from these curious simplifications in higher-dimensional CFTs.\footnote{For a connection to gravitational shockwaves, see \cite{Fitzpatrick:2019efk}, where a further analysis based on a spherical black hole can be found. Note that a planar black hole limit suppresses derivatives inserted into stress tensors, resulting in just one leading-twist $T^n$ primary at each $n$. The differential equations discussed in this work, as we will verify, apply to the holographic computation with a spherical black hole where an {\it infinite} number of leading-twist $T^n$ primaries contribute at each $n\geq 2$.} For concreteness we consider $d=4$, where the corresponding special values are $\Delta =-1,-2,-3$.  

We have excluded the $d=2$ case from \eqref{specialDelta}. The reason is as follows: it appears that the corresponding computation in AdS$_3$ is somewhat ``too simple" to detect any special conformal dimensions, and the closed form of the Virasoro vacuum block at large $c$ with a general conformal dimension can be obtained directly. In the $d=2$ case, where we denote $\Delta \equiv 2 h_L$, we  assume that the corresponding special value is $h_L= -\frac{1}{2}$. This value, as we will review, can be derived using the Virasoro algebra. 

By adding non-minimally coupled bulk interactions, some explicit corrections to the leading-twist OPEs in $d=4$ were found in \cite{Fitzpatrick:2020yjb}, showing that the near-lightcone correlator at large $C_T$ can depend on the higher-spin gap scale.
In this work, we observe that the corrections due to the quartic derivative, non-minimally coupled interactions between the graviton and the light bulk scalar computed in \cite{Fitzpatrick:2020yjb} vanish precisely at the values $\Delta= - 1, -2, -3$.  These are the required values in the $d=4$ linear differential equations we will discuss below.  It is thus tempting to speculate that the differential equations considered in this work might apply to a broader class of conformal theories.  

The resummations of the leading-twist double- and triple-stress tensors were obtained in \cite{Kulaxizi:2019tkd, Karlsson:2019dbd, Karlsson:2020ghx}. These results correspond to a holographic computation utilizing a spherical black hole.
It was shown that the heavy-light correlator in $d=4$ (and similarly in other even dimensions) has a general structure
\begin{align}
\label{AP}
F(z,\bar z)=  1+
 \sum_{n=1}^{\infty}   \Big[ \sum_{i_p=1}   A_{i_1 \cdots i_n}(\Delta) ~ f_{i_1}(z) \cdots f_{i_n}(z) \Big]  (\mu \bar z)^{n},  ~~~  \sum_{p=1}^{n} i_p =  3n  
\end{align}   
where $p=1,2, \cdots, n$, and  $f_a(z) \equiv z^a~_2F_1(a,a,2a,z)$ involves the Gaussian hypergeometric function.\footnote{Our $(z, \bar z)$ corresponds to $(1-z, 1-\bar z)$ in \cite{Kulaxizi:2019tkd, Karlsson:2019dbd, Karlsson:2020ghx}.}   
The coefficients $A_{i_1 \cdots i_n}(\Delta)$ depend on the number of exchanged stress tensors.  
The indices $i_p \geq 1$ are integers subjected to the constraint above.
 Unrestricted to the special values \eqref{specialDelta},  the  expression for the $n= 2$ case 
was found in \cite{Kulaxizi:2019tkd} based on the multi-stress tensor OPEs computed from holography.  
Subsequently, both $n=2$ and $n= 3$ expressions were computed in \cite{Karlsson:2019dbd, Karlsson:2020ghx} by employing the lightcone bootstrap, treating \eqref{AP} as an ansatz. Certain similarity between such a $d=4$ structure and the $d=2$ ${\cal W}_3$ vacuum blocks was  discussed in \cite{Huang:2021hye, Karlsson:2021mgg}. On the other hand, the Lorentzian inversion formula \cite{Caron-Huot:2017vep, Simmons-Duffin:2017nub} also allows one to compute the OPE coefficients of multi-stress tensor exchanges at large $C_T$ \cite{Li:2019zba, Li:2020dqm}.  
To compute the OPEs, it appears that, starting from triple-stress tensor exchanges, the ansatz \eqref{AP} combined with the inversion formula or crossing symmetry is more efficient than only using the inversion formula, as remarked in \cite{Karlsson:2020ghx}.\footnote{Further advancements have been made. For instance, a closed-form expression of the $d=4$ heavy-light correlator at large $\Delta$ was obtained in \cite{Parnachev:2020fna} based on a holographic computation with a planar black hole.  The thermalization of multi-stress tensor operators was discussed in \cite{Karlsson:2021duj}. Another recent study \cite{Dodelson:2022yvn} adopted a formal connection to the Nekrasov-Shatashvili partition function \cite{Nekrasov:2009rc} of a supersymmetric gauge theory to compute the $d=4$ thermal scalar two-point correlator through holography.}  In this work, we are interested in an approach via BPZ-type linear differential equations for analyzing the near-lightcone correlators at large $C_T$ in $d>2$. 

Let us comment on the required negative values of $\Delta$. Recall that many remarkable results in $d=2$ CFTs are consequences of shortening conditions of the conformal algebra at certain values of conformal dimensions and the central charge. At large central charge, the corresponding dimensions of the light scalars take {\it non-unitary} values, $i.e.$, negative integers. This may appear to be a serious limitation, but, as emphasized in \cite{Chen:2016cms}, since the conformal blocks are analytic functions of their defining quantum numbers, the Virasoro blocks with general conformal dimensions can be obtained by an analytical continuation of the correlators involving degenerate operators. 
We may take the same perspective in the $d=4$ case.  The focus of this paper, however, is to identify and analyze a set of differential equations in $d=4$ and we will not delve into an additional investigation of the analytical continuation. 

In contrast to the $d=2$ case where the large-$c$ BPZ equations are ordinary differential equations, the $d=4$ equations discussed below are $\it partial$ differential equations. 
This is not surprising, given that the correlator in the near-lightcone regime depends on the coordinates $z$ and $\bar z$ in a conformal frame.   
While solving the $d=4$ equations is more challenging than the $d=2$ case, we find them straightforward to solve in a small $\bar z$ expansion with appropriate boundary conditions.  The resulting solutions, valid for $\Delta=-1,-2,-3$, are consistent with those previously derived via holography and the lightcone bootstrap. These involve the expressions that resum an infinite number of leading-twist multi-stress tensors. These new differential equations also enable us to make predictions about the near-lightcone correlator. 

We now provide a summary of the main equations. 

\vspace{0.2cm}

\noindent \underline
{{\it Two dimensions}}:  

To compare to the $d=4$ case, it is useful to first present the corresponding differential equation in $d=2$. 
 The level-two BPZ null-state equation \cite{BELAVIN1984333}, when applied to the heavy-light correlator at large central charge, can be expressed as
\begin{align}
\label{2dBPZ}
Q''(z)+ 6 \eta {Q(z)\over (1-z)^2}=0   \ ,  ~~~~~ Q(z) \equiv  z F(z)  
\end{align} 
which is valid at $h_{L}= {\Delta \over 2 }= -  {1 \over 2 }+ {\cal O}({1\over c})$. 
Here the function $F$ corresponds to \eqref{defineF}, but the $\z$ dependence is neglected as the focus here is on the holomorphic correlator.\footnote{More precisely, in $d=2$ we write 
\begin{align}
F(z)={\langle {\cal O}_H (\infty) {\cal O}_L(z) {\cal O}_L(0)  {\cal O}_H(1)\rangle \over   \langle {\cal O}_H(\infty)  {\cal O}_H (1) \rangle \langle {\cal O}_L(z) {\cal O}_L (0) \rangle}\Big|_{T^n} 
= \sum^{\infty}_{n=0} G_n (\Delta, z) \eta^n \ .
 \end{align} Instead of using the notations $\mu$ and $C_T$, we adopt $\eta= {h_H\over c}$ in $d=2$  where $h_H$ represents the weight of the heavy scalars. The central charge $c$ can be defined via  $\langle T(z)T(0) \rangle = {c\over 2 z^4}$.} 
The function $Q(z)$ in the $d=2$ case satisfies a slightly more concise equation compared with the equation of $F(z)$. This type of simplification will be particularly useful in identifying and representing the linear differential equations in $d=4$. In the $d=2$ case, the factor $z$ in the relation $Q= z F$ corresponds to the scalar two-point function, ${z^{-2 h_L}}=z$. The relation between functions $Q$ and $F$ will vary slightly in $d=4$. 

It is straightforward to solve the equation \eqref{2dBPZ}. The solution  
\be
\label{2dresult}
F(z)  = {1\over z }  { (1-z)^{\frac{1}{2}-\frac{1}{2} \sqrt{1-24\eta }}  \left(1- (1-z)^{\sqrt{1-24\eta }}\right) \over \sqrt{1-24\eta }}
\ee corresponds to the large-$c$ heavy-light Virasoro vacuum block when $h_{L}= -  {1 \over 2 }$ \cite{Fitzpatrick:2014vua, Fitzpatrick:2015zha}. One may extract  specific contributions from exchanged $T^n$ operators via the expansion $F(z)= \sum_{n=0} G_n(z) \eta^n$.  (Incorporating an ${\cal O}({1\over c})$ correction requires the level-3 BPZ equation. See a related  discussion in \cite{Chen:2016cms}.)

\vspace{0.2cm}

\noindent  \underline{{\it Four dimensions}}:

To take the story further,  we hypothesize that at the special values \eqref{specialDelta}, certain $d=4$ generalized primaries at large $C_T$ have null descendants, thereby necessitating their correlators to obey linear differential equations resembling \eqref{2dBPZ}.  

As mentioned, the parameter $\mu$ can be absorbed into $\bar z$ in the near-lightcone expansion. We perform such a rescaling to remove the explicit dependence on $\mu$ in the following $d=4$ expressions. 

We introduce 
\begin{align}
\label{QF}
Q_{\Delta^{*}}(z,\bar z) \equiv   {z\over (z \bar z)^{\Delta}}  F(z,\bar z)\big|_{\Delta=\Delta^*} \ , ~~~~~ \Delta^*= -1, -2, -3 \ ,
\end{align}
where $F(z,\z)$ is defined in  \eqref{defineF}. We propose the following linear differential equations to be the $d=4$ generalizations to the BPZ equation $\eqref{2dBPZ}$ at large central charge:
\begin{align}
\label{eqm1}  
&Q_{-1}^{(3,1)}+  { Q_{-1} \over (1-z)^3}=0 \ ,    \\
 \label{eqm2}
&Q_{-2}^{(4,1)}+6 \frac{ Q_{-2}^{(1,0)}}{(1-z)^3}+ 9 \frac{ Q_{-2}}{(1-z)^4}= 0 \ ,  \\
 \label{eqm3}
&Q_{-3}^{(5,1)}+21 \frac{Q_{-3}^{(2,0)}}{(1-z)^3}+ 63 \frac{Q_{-3}^{(1,0)}}{(1-z)^4}+72 \frac{Q_{-3}}{(1-z)^5} = 0 
\end{align} 
where $Q_{\Delta^{*}}^{(m,n)} = \partial^{m}_z \partial^n_{\bar z}Q_{\Delta^{*}}  (z,\bar z)$. 

Some preliminary remarks are in order:

1.  
Crossing symmetry, due to the presence of two identical light scalars, imposes a strong constraint on the possible forms of the equations. By translating these $d=4$ equations to the ones using the original four-point function $F(z,\z)$, the resulting equations transform covariantly under $z \to {z \over z-1}$,  $\z \to {\z \over \z-1} \approx  -\z$, where a small $\z$ is assumed.  Note that, however, crossing symmetry along is insufficient to fully determine the equations.  

2. 
The equation \eqref{eqm1} is formulated by drawing upon the BPZ equation in the form of the Schr\"{o}dinger equation \eqref{2dBPZ}, but instead of a second-order derivative we consider a higher-derivative structure.  
Assuming the equation takes the form $Q^{(\alpha,\beta)}+ \k {Q/(1-z)^{\gamma}}=0$, the consistency with the $d=4$ universal single-stress tensor exchange in the near-lightcone expansion fixes $\alpha, \beta, \gamma$ and $\kappa$ (note that the relative coefficient, $\kappa$, depends on the normalization of the single-stress tensor exchange which has a canonical convention). In fact, with the $Q^{(3,1)}$ structure, crossing symmetry fixes the power of $(1-z)$. The resemblance between the BPZ equation  \eqref{2dBPZ} and the $d=4$ equation \eqref{eqm1}, as we will demonstrate below,  leads to analogous recursive integral formulas in both $d=2$ and $d=4$, which provide a systematic and efficient method for computing the correlators. 

3.  
The equation form \eqref{eqm2} is obtained by applying one additional $z$-derivative to \eqref{eqm1}, with different  constant coefficients.  Likewise, the form \eqref{eqm3} is derived by applying two more $z$-derivatives to \eqref{eqm1}. Treating the universal single-stress tensor exchange as an input enables completely fixing the constant coefficients -- the unexpected bonus is that the higher-order solutions in the near-lightcone expansion, which correspond to the exchanges of  multi-stress tensors, automatically reproduce the previously computed results via holography and the lightcone bootstrap.    
We find that this ``derivation" of the equations does {\it not} apply to other values of $\Delta$, in agreement with the assumed conditions of \eqref{specialDelta}.\footnote{A source of intuition comes from $d=2$, where the large-$c$ correlator with negative dimensions other than $h_L=-{1\over 2}$ satisfies higher-order BPZ equations. However, in the $d=4$ case, there exist three special values of $\Delta$ in the {\it leading} large $C_T$ limit, whereas in $d=2$, there is only one value.}
We emphasize that these equations are not derived by fitting with the known results of  {\it multi}-stress tensor contributions (although in a first-principle approach the single-stress tensor exchange contribution should also be an output). 
 As we get more out of these equations than was originally put into them, in this paper we explore their consequences.  Further examination of these equations is certainly warranted.

4. 
An interesting property of these $d=4$ ``$Q$-equations" \eqref{eqm1}$-$\eqref{eqm3}, which are linear partial differential equations, is that the coefficients do not depend on $\bar z$.  Moreover, the derivative of $\z$ appears with an order no greater than one.  Expressing these equations in terms of $F(z, \bar{z})$ results in longer equations with coefficients that explicitly depend on $\bar{z}$.\footnote{In the abstract, we let $Q = u$, $z = 1-x$, $\bar z = - y$ as a way to represent a $d=4$ partial differential equation, and we drop the $\eta$ dependence in the $d=2$ equation as a schematic form.} To our knowledge,  
there is no literature linking this type of partial differential equations to any known physical phenomenon.

5. 
The BPZ equations can be derived using the Virasoro algebra.  We do not have a similar derivation of these $d=4$ equations. First-principle derivations, based on a symmetry or not, of these equations are left for the future.  We hope that this work may stimulate one to think of possible symmetry enhancements in $d>2$ CFTs. 

\vspace{0.15cm}

This article is organized as follows. Section 2 provides a brief review of the $d=2$ case and presents a recursive integral formula for the heavy-light correlator. In Section 3, we examine the $d=4$ differential equations and verify that their solutions in the near-lightcone expansion are consistent with the results obtained via holography and the lightcone bootstrap. Corresponding integral formulas are also obtained. Using these formulas, we present some predictions of the $d=4$ equations. We conclude with additional remarks in Section 4.

\subsection*{2. $d=2$ BPZ at large $c$} 

One can derive the null-state equation \eqref{2dBPZ} starting with the Virasoro algebra 
\begin{align}
\label{Virasoro}
[L_m, L_n]= (m-n) L_{m+n} +  {c\over 12} m (m^2-1) \delta_{m+n,0} \ .
\end{align}  
We are interested in Virasoro descendants with a zero norm. 
As is well-known, the simplest, non-trivial null state is the level-2 descendant, which can be represented as 
\begin{align}
\label{null2}
|\chi\> = \big( L^2_{-1}+ b^2 L_{-2} \big) | h \>  \ , ~~~~~~~ c = 1+ 6 \Big( b + {1\over b }\Big)^2  \ . 
\end{align} 
The vanishing determinant of the level-2 Gram matrix requires the weight $h$ to be 
\begin{align}
\label{2dgh}
h= -{1\over 2} - {3 \over 4 b^{2} } \ .
\end{align}  
A null state is orthogonal to any state in the theory. This property translates, using the mode operator $L_{-m}$ together with the Ward identities, into the level-2 BPZ differential equation \cite{BELAVIN1984333}: 
\begin{align}
\label{2dgBPZ}
\Bigg( \partial_z^2 + \Big( \frac{b^{-2}}{1-z}+ \frac{2+2 b^{-2}}{z} \Big) \partial_z  + \frac{ h_H b^{-2}}{(1-z)^2} \Bigg) F(z) = 0  \ .
\end{align}
Here we focus on the heavy-light correlator $F(z)= \frac{\< {\cal O}_H(\infty) {\cal O}_H(1) {\cal O}_{h}(z) {\cal O}_{h}(0) \>}{\< {\cal O}_H(\infty) {\cal O}_H(1) \>\< {\cal O}_{h}(z) {\cal O}_{h}(0) \>}$ with the dimension of the light scalars given by \eqref{2dgh}. 
By taking the dimensions of the heavy scalars to be of order $c$, $h_H \sim {\cal O} (c)$, the null-state equation at large $c$  reduces to 
\begin{align}
\label{2dF}
\Big( \partial^2_z + {2\over z}  \partial_z \Big) F(z)+ { 6  \eta\over (1-z)^2}   F(z)=0  \ , ~~~~~ \eta= {h_H\over c} \ .
\end{align} 
Note that $h \to -{1\over 2}$ at large $c$, where the parameter $b \to \infty$.   This equation transforms covariantly under the exchange of the two identical light scalars, $i.e.$, $z \to {z \over z-1}$.\footnote{We note that crossing symmetry fixes the ratio $1:2$ in the derivative part of the large-$c$ BPZ equation \eqref{2dF}, but is not enough to fully determine the equation.} The Schr\"{o}dinger-type equation \eqref{2dBPZ} is obtained by using $Q(z)= z F(z)$.

The equation \eqref{2dF} has the following general solution:
\begin{align}
F(z)= {1\over z} \left(c_1+  c_2 (1-z)^{\sqrt{1-24 \eta }}\right) (1-z)^{\frac{1}{2}-\frac{1}{2} \sqrt{1-24 \eta }}  \  .
\end{align}
The coefficients $c_1$ and $c_1$ can be fixed by a small $z$ expansion:
\begin{align}
\lim_{z\to 0}F(z) = \frac{c_1+c_2}{z}+\left(\frac{1}{2} (c_1+c_2) \left(\sqrt{1-24 \eta }-1\right)-c_2 \sqrt{1-24 \eta }\right)+{\cal O}\left(z\right) \ .
\end{align} The regularity at $z \to 0$ requires $c_1= - c_2$. The remaining coefficient can be fixed by  the 
normalization $F(0)=1$, giving $c_2= -\frac{1}{\sqrt{1-24 \eta}}$. The result, presented in \eqref{2dresult}, is the large-$c$ heavy-light Virasoro vacuum block at $h= - {1\over 2}$.   The same consideration applies to the anti-holomorphic part of the correlator.  

Although the closed-form solution to the $d=2$ equation \eqref{2dF} can be obtained straightforwardly, 
performing a perturbative analysis in a $\mu$-expansion will be useful for the discussions in $d=4$. This will enable us to observe common patterns among the $d=2$ and $d=4$ perturbative solutions.

Consider a near-lightcone expansion 
\begin{align}
F(z)= 1+ \eta  G_1(z)+\eta ^2 G_2(z)+\eta ^3 G_3(z)+ {\cal O} (\eta^4)
\end{align} 
which has incorporated the normalization $F(0)=G_0(0)=1$. 
The BPZ equation in such an expansion exhibits a recursive pattern:
{\small \begin{align}
0=&\left(z G_1''(z)+2 G_1'(z)+\frac{6 z}{(1-z)^2}\right) \eta \\
&+\left(z G_2''(z)+2 G_2'(z)+\frac{6 z G_1(z)}{(1-z)^2}\right) \eta^2 +\left(z G_3''(z)+2 G_3'(z)+\frac{6 z G_2(z)}{(1-z)^2}\right) \eta ^3 +{\cal O} (\eta^4) \nn \ .
\end{align}}The general solution of $G_1$ is
\begin{align}
G_1=  {1\over z} \Big(c_1 +c_2 z-6 (z-2) \ln (1-z)\Big)
\end{align} 
where constants $c_1$ and $c_2$ can be fixed by the $z\to 0$ behavior. Requiring $G_1(0)=0$ gives $c_1=0$, $c_2= 12$, yielding the correct result for the single-stress tensor exchange near the lightcone
\begin{align}
G_1= - { 1 \over z} \Big(6 (z-2) \ln (1-z)- 12 z \Big) = - f_2 
\end{align} 
where $f_a \equiv f_a(z) = z^a~_2F_1(a,a,2a,z)$. Similarly, the higher-order solutions can be obtained:
{\allowdisplaybreaks
\begin{align}
G_2& =  {18 \over z} \Big(z \ln ^2(1-z)- 6 (z-2) \ln (1-z)+12z\Big) \nn\\
&= -\frac{6}{5} f_1 f_3+\frac{3}{2} f_2 f_2 \ , \\
G_3& 
=  -\frac{36 }{z} \Big((z-2) \ln^3(1-z)-12 z \ln ^2(1-z)+60 (z-2) \ln (1-z)-120 z\Big) \nn\\
&= -\frac{54}{35} f_1 f_1 f_4 +4 f_1 f_2  f_3-\frac{5}{2} f_2 f_2 f_2\ .
\end{align}}We note in passing that a pattern similar to \eqref{AP} appears: $G_n \sim \sum_{i_p} f_{i_1}\dots f_{i_p}$ with $\sum_{p=1} i_p= 2n$. This pattern in $d=2$ has an origin in the Virasoro algebra \eqref{Virasoro}.  

Let us introduce an integral formula to organize the perturbative solutions. By plugging the expansion
\begin{align}
Q(z)= \sum_{n=0} R_n(z) \eta^n   \ , ~~~ R_0(z)= z
\end{align}
 into the equation  \eqref{2dBPZ} and matching the power of $\eta$, we deduce that 
\begin{align}
\label{int2d}
R_{n+1}(z)= - 6 \int^z_0 ds_2 \int^{s_2}_0 ds_1  ~\frac{R_n(s_1)}{(1-s_1)^2} \ .
\end{align}   
The double integrals arise from the second-order differential equation. 
This formula reproduces the $G_1, G_2, G_3$ results (using $R_{n}= z G_n$) and presents a streamlined approach to determine the correlator at higher orders. While this approach is somewhat redundant in $d=2$, the similar formulas in $d=4$, which we will present below, provide an efficient way to analyze the near-lightcone structure of the correlator.

These $d=2$ correlator results can be obtained using holography by considering a bulk scalar two-point function with a BTZ black hole in AdS$_3$.  In higher dimensions, a similar gravity computation is also possible, as mentioned in the introduction.  The motivation of this work is to find a way to calculate $d>2$ correlators at large central charge by utilizing a method similar to the $d=2$ approach discussed in this section.  In higher dimensions, we will focus on the correlator structure near the lightcone.  

\subsection*{3. $d=4$ equations at large $C_T$}  

In this section, we examine the $d=4$ linear differential equations \eqref{eqm1}$-$\eqref{eqm3}, which are proposed as the generalizations to the $d=2$ equation \eqref{2dBPZ} or, equivalently,  \eqref{2dF}.   

\vspace{0.25cm}

\noindent {\it \small 3.1 $\Delta= -1$ case}

It may be illuminating to first take a look at the equation \eqref{eqm1} in a different form, adopting the function $F(z,\z)$ with the parameter $\mu$ restored:
\begin{align}
\label{m1Feq}
\Big(1+ \z \partial_{\z} \Big) \Big(  \partial^3_z+ {6 \over z} \partial^2_z+{6\over z^2} \partial_z\Big)F (z,\z)+ {\mu \z \over (1-z)^3} F(z,\z) = 0 \ .
\end{align} 

Compared to the $d=2$ equation \eqref{2dF}, a notable difference is that this $d=4$ equation includes both $z$- and $\z$-derivatives. A factorization between the $z$-derivative part and the $\z$-derivative part of the equation can be observed.  The equation transforms covariantly under $z \to {z \over z-1}$,  $\z \to - \z$ (in the near-lightcone limit).\footnote{Here $z$ and $\z$ are two independent real variables and we explore the near-lightcone limit, $\z \to 0$. In Euclidean signature, $z$ is a complex number and $\z$ is its conjugate.} 
Crossing symmetry is expected as two light scalars are identical. Similar to the $d=2$ case,  the crossing relation fixes the ratio $1:6:6$ in the bracket of the equation \eqref{m1Feq} involving $z$-derivatives, but is insufficient to uniquely determine the full equation.  It would be interesting to construct certain mode operators, possibly with regularization in the near-lightcone regime, as well as a null state to provide a derivation of the $d=4$ equation.   

The partial differential equation \eqref{m1Feq} is non-trivial to solve directly.  Here we adopt an approximate analysis.  
Consider an expansion
\begin{align}
\label{defineG}
F(z,\z)= 1+  G_1(z) \z+  G_2(z) \z^2 + G_3(z)  \z^3+ {\cal O} (\z^4)
\end{align} 
where we have fixed a normalization and absorbed the parameter $\mu$ into $\z$.  This near-lightcone expansion \eqref{defineG} will be used for all cases with $\Delta=-1,-2,-3$.

The differential equation \eqref{m1Feq} leads to  
{\small
\begin{align}
\label{4dm1pereq}
0=&\left(2 z^2 G_1^{(3)}(z)+12 z G_1''(z)+12 G_1'(z)+\frac{z^2}{(1-z)^3}\right)\z \nn\\
&~~+ \left(3 z^2 G_2^{(3)}(z)+18 z G_2''(z) +18 G_2'(z) +\frac{z^2 G_1(z)}{(1-z)^3} \right) \z^2  \\
&~~~~~~~+\left(4 z^2 G_3^{(3)}(z)+ 24  z G_3''(z)+24 G_3'(z) +\frac{z^2 G_2(z)}{(1-z)^3}\right)\z^3 + {\cal O}\left(\z^4\right) \ . \nn
\end{align}}The general solution of $G_1$ is 
\begin{align}
G_1= {1\over 4 z^2 } \Big( c_1 + c_2 z+ c_3 z^2 +\big(6+(z-6) z\big) \ln (1-z)\Big)  \ .
\end{align} We can consider a small $z$ limit and require $G_1(0)=0$ to determine that $(c_1, c_2, c_3) = (0 ,6, -3)$.
The result is the universal single-stress tensor exchange contribution near the lightcone:  
\begin{align}
G_1={1\over 4 z^2 }  \Big(\big(6+(z-6) z\big) \ln (1-z)-3 (z-2) z\Big) = - { f_3\over 120} \ . 
\end{align} 

Proceeding to solve for $G_2$ and $G_3$, which correspond to double- and triple-stress tensor exchanges, respectively,  we find 
{\allowdisplaybreaks 
\begin{align}
\label{G2m2}
G_2&= {1\over 48 z^2} \Big[\big(z (z+6)-6\big) \ln ^2(1-z)-9 (z-2) z \ln (1-z) + 24 z^2\Big]\nn\\
&=  -\frac{f_2 f_4 }{4480}  +\frac{ f_3 f_3}{4320}  \ , \\
G_3&= {1\over 1152 z^2} \Big[(z (z-18) +18) \ln^3(1-z) -18 (z-2) z \ln^2 (1-z)  \nn\\
\label{G3m2}
&~~~~~~~~~~~~~~~~~+ \big( 3 (41 z-210) z +630 \big) \ln (1-z) -315 (z-2) z \Big]\nn\\
&=  -{5 f_1 f_2 f_6 \over 6386688} + {f_1 f_3 f_5 \over 1209600}  -{f_2 f_2  f_5\over 2903040}  + { f_2 f_3 f_4\over 483840} - {11  f_3 f_3 f_3\over 6220800} \ . 
\end{align}}In this $\Delta=-1$ case, a possible $f_1 f_5$ term in \eqref{G2m2} and an $f_1 f_1 f_7$ term in \eqref{G3m2} both have a vanishing coefficient.\footnote{The expressions written in terms of hypergeometric functions are not unique due to relations among them. For instance, $9 f_2 f_4 - {28\over 3}  f_3 f_3= {140} (f_1 f_3  - f_2 f_2)$.\label{fo8}}
The $G_2$ result obtained here using the differential equation is in agreement with that obtained in \cite{Kulaxizi:2019tkd}  based on multi-stress tensor OPEs computed from holography.  The $G_3$ result is in agreement with \cite{Karlsson:2019dbd, Karlsson:2020ghx} which employ the lightcone bootstrap.

To compute higher-order results, let us derive an integral formula using the equation \eqref{eqm1}.  
The derivation is essentially identical to the $d=2$ case. Here we take  
\begin{align}
\label{Rm1d}
 Q(z,\z)= \sum_{n=0} R_n(z) \z^{n+1} \ , ~~~ R_0(z)= z^2 \ ,
\end{align}    where the subscript of $Q$ is suppressed. 
 By matching the power of $\z$, we obtain
\begin{align}
\label{Intm1}
R_{n+1}(z)=  {-1\over n+2} \int^z_0 ds_3 \int^{s_3}_0 ds_2 \int^{s_2}_0 ds_1  ~\frac{R_n(s_1)}{(1-s_1)^3} \  .
\end{align} 
This formula reproduces the $G_1, G_2, G_3$ results (recall $ R_n= z^2 G_n$ in this case) 
and provides an efficient way to compute higher-order contributions.\footnote{For instance, {\allowdisplaybreaks 
{\small \begin{align}
G_4&= {1\over 46080 z^2} \Big(\left(z^2+30 z-30\right) \ln^4(1-z) -30 (z-2) z \ln^3(1-z)\nn\\
&~~~ +15 \left(25 z^2+78 z-78\right) \ln^2(1-z)-2295 (z-2) z \ln (1-z)+  5760 z^2\Big)\\
&=  -{f_1 f_2 f_2 f_7 \over 1014773760}  -{17 f_1 f_1 f_3 f_7\over 1353031680} + {5 f_2 f_2 f_2 f_6\over 204374016}  + {643 f_1 f_2 f_3 f_6 \over 224811417600} \nn\\
& ~~~ + {433 f_1 f_1 f_4 f_6 \over 58284441600} + {3 f_1 f_1 f_5 f_5 \over 551936000}  -{7 f_2 f_2 f_3 f_5\over 182476800}- {7 f_1 f_3 f_3 f_5 \over 3649536000}  +  {7 f_3 f_3 f_3 f_3\over 513216000} 
\end{align}}}The expression written as hypergeometric functions allow varying forms.}

Let us test the equation \eqref{eqm1} further. By utilizing the formula \eqref{Intm1} to compute higher-order terms, we may focus on the leading small-$z$ piece of each function $G_n$. We observe a pattern that allows for the resummation over $n$:
\begin{align}
\label{m1resumsmallz}
\sum_{n=0}^{\infty}\frac{2 \left(-z^3 \z \right)^n}{\Gamma (n+2) \Gamma (3 n+3)} 
=~ _0F_3\left(\Big\{\frac{4}{3},\frac{5}{3},2\Big\} ,-\frac{z^3 \z}{27} \right) \ . 
\end{align} 
Note that this is an expression for $F(z,\z)$ where the identity contribution due to the $n=0$ term is included in the resummation. We observe that \eqref{m1resumsmallz} agrees with the corresponding correlator result obtained holographically by solving a bulk scalar equation of motion with a planar black hole in AdS$_5$ \cite{Fitzpatrick:2019zqz}.\footnote{Specifically, we are referring to Eq. (4.27) in that paper.} This result corresponds to a resummation of {\it all} leading-twist multi-stress tensors without derivatives inserted. 

Another related observation is that the expression $(z^2 \z)~ _0F_3\left(\Big\{\frac{4}{3},\frac{5}{3},2\Big\} ,-\frac{z^3 \z}{27} \right)$,  which was first obtained via holography, is an  {\it exact} solution to the partial differential equation $Q^{(3,1)}(z,\z)+ Q(z,\z)=0$. This equation corresponds to \eqref{eqm1} with a replacement of the factor $(1-z)^3 \to 1$.

The differential equation \eqref{eqm1} (or equivalently \eqref{m1Feq}) allows us to make predictions on the correlator structure  beyond the leading small-$z$ (or planar black hole) limit. Specifically, by using higher-order solutions generated from the formula \eqref{Intm1} and collecting both sub-leading and sub-sub-leading small-$z$ contributions (at each order of $\z$), we can identify the patterns allowing for the resummation.  The corresponding results for the function $F(z,\z)$ are given by 
{\allowdisplaybreaks {\begin{align}
&\sum_{n=1}^{\infty} \frac{3n }{\Gamma (n+2) \Gamma (3 n+3)}  \left(-z^3 \z \right)^n z= -\frac{z^4  \z}{80} \,   _0F_3 \left(\Big\{\frac{7}{3},\frac{8}{3},3 \Big\}, -\frac{z^3 \z }{27} \right) \ , \\
&\sum_{n=1}^{\infty} \frac{\pi  3^{-3 n-\frac{5}{2}} (9 n^2+17 n +6) }{2 \Gamma (n)  \Gamma \left(n+\frac{5}{3}\right) \Gamma (n+2)  \Gamma \left(n+\frac{7}{3}\right)}  \left(-z^3 \z \right)^n z^2\nn\\
&=- {z^5 \z \over 2240} \Big[6 ~\, _0F_3\left(\Big\{\frac{8}{3},3,\frac{10}{3}\Big\},-\frac{z^3 \z}{27} \right)
+17~\,  _1F_4\left(\Big\{2\Big\}, \Big\{1,\frac{8}{3},3,\frac{10}{3}\Big\},-\frac{z^3 \z}{27}\right)\nn\\
&~~~~~~~~~~~~~~~~~~~~~~~~~ +9 ~\, _2F_5\left(\Big\{2,2\Big\}, \Big\{1,1,\frac{8}{3},3,\frac{10}{3}\Big\},-\frac{z^3 \z}{27} \right) \Big]  \ . 
\end{align}}}More higher-order results can be worked out. Obtaining these new results through holography may be a non-trivial task as it requires a {\it spherical} black hole. 


\vspace{0.25cm}

\noindent {\it \small {3.2 $\Delta= -2$ case}}

The equation \eqref{eqm2} written in terms of the function $F(z,\z)$ is given by
{\allowdisplaybreaks
\begin{align}
&0=
\Big( 2+\z \partial_{\z}\Big) \left( \partial_z^4 + {12\over z}  \partial_z^3 +{36\over z^2}  \partial_z^2 +{24 \over z^3}  \partial_z \right) F(z,\z)  \nn\\
&~~~~~~~~~~~~~~~~~~~~~~~~~
+\frac{6 \mu \z }{(1-z)^3}\Big( \partial_z+ \frac{3 (2-z)}{2  z(1-z) } \Big) F(z,\z) \ .
\end{align}}In this expression we have rescaled $\z$ to make the parameter $\mu$ explicit.  This equation is also consistent with crossing symmetry.  Solving this equation via the near-lightcone expansion, using the same boundary conditions discussed above, is straightforward. 

To derive an integral formula, we adopt the equation \eqref{eqm2} and write
\begin{align}
\label{Rm2d}
 Q(z,\z)= \sum_{n=0} R_n(z) \z^{n+2} \ , ~~~ R_0(z)= z^3 \ . 
\end{align}   
(The subscript of $Q$ is suppressed and $\mu$ is absorbed into $\z$.) We find
\begin{align}
\label{Intm2}
R_{n+1}(z)=  {-1\over n+3} \int^z_0 ds_4 \int^{s_4}_0 ds_3 \int^{s_3}_0 ds_2 \int^{s_2}_0 ds_1  ~ 
\Big[ \frac{ 9 R_n(s_1)}{(1-s_1)^4}+ \frac{ 6 R'_n(s_1)}{(1-s_1)^3}  \Big] \ .
\end{align}
Using $R_n= z^3 G_n$, where $G_n$ is defined via  \eqref{defineG}, in this $\Delta=-2$ case we have verified that the formula \eqref{Intm2} reproduces the universal $G_1$, and the known $G_2, G_3$ expressions obtained in \cite{Kulaxizi:2019tkd, Karlsson:2019dbd, Karlsson:2020ghx}. Higher-order contributions can be  obtained readily. 

Although we shall not provide an exhaustive list of higher-order solutions, let us mention that the resummation over $n$, similar to those in the $\Delta=-1$ case, can be obtained in a small $z$ limit.  At the first three orders in the $z\to 0$ expansion, we have (for the function $F(z,\z)$ in the near-lightcone limit)
{\allowdisplaybreaks\small \begin{align}
\label{planar2}
&\sum_{n=0}^{\infty} \frac{2^{n+2} 3^{n+1} }{\Gamma (n+3) \Gamma (3 n+4)}  \left(-z^3 \z \right)^n
= {1\over {z^3 \z}} \Big[2-2 \, _0F_3\left(\Big\{\frac{1}{3},\frac{2}{3},2\Big\},-\frac{2 z^3 \z}{9} \right) \Big] \ , \\
&\sum_{n=1}^{\infty}  \frac{2^{n+1} 3^{n+2} n }{\Gamma (n+3) \Gamma (3 n+4)}  \left(-z^3 \z\right)^n z
=- {3\over  z^2 \z}
\Big[1+ \, _0F_3\left(\Big\{\frac{1}{3},\frac{2}{3},1\Big\}, -\frac{2 z^3 \z}{9}\right)\nn\\
&~~~~~~~~~~~~~~~~~~~~~~~~~~~~~~~~~~~~~~~~~~~~~~~~~~~~~~~~
 -2 \, _0F_3\left(\Big\{\frac{1}{3},\frac{2}{3},2\Big\}, -\frac{2 z^3 \z}{9} \right)\Big] \ ,  \\
&\sum_{n=1}^{\infty} \frac{2^{n-1} 3^{n+2} n \big(3 n (9 n^2+32n +34)+31\big) }{\Gamma (n+3) \Gamma (3 n+6)}  \left(-z^3 \z \right)^n z^2\nn\\
&= \frac{1}{8960 z \z}
\Bigg[\frac{54}{55} z^9 \z^3 \, _0F_3\left(\Big\{4,\frac{13}{3},\frac{14}{3}\Big\},-\frac{2 z^3 \z}{9} \right)+84 \Bigg(12 z^3 \z \, _0F_3\left(\Big\{2,\frac{7}{3},\frac{8}{3}\Big\},-\frac{2 z^3 \z}{9}\right)\nn\\
&~~~~~~~~~~~~~~~~+320 ~\, _0F_3\left(\Big\{1,\frac{4}{3},\frac{5}{3}\Big\},-\frac{2 z^3 \z}{9}\right)-400 ~ \, _0F_3\left(\Big\{\frac{4}{3},\frac{5}{3},2\Big\}, -\frac{2 z^3 \z}{9} \right)\nn\\
&~~~~~~~~~~~~~~~-3 z^6 \z^2 \, _0F_3\left(\Big\{3,\frac{10}{3},\frac{11}{3}\Big\},-\frac{2 z^3 \z}{9}\right)+80\Bigg)\Bigg] \ .
\end{align}}The leading-order result \eqref{planar2}, which includes the identity operator, is in perfect agreement with the holographic calculation with a black brane.\footnote{See \cite{Fitzpatrick:2019zqz}, Eq. (4.25), for verification.} The remaining two expressions that go beyond the leading small-$z$ limit are predictions of \eqref{eqm2}. 

\vspace{0.25cm}

\noindent {\it \small {3.3 $\Delta= -3$ case} }

The equation \eqref{eqm3} expressed in terms of $F(z,\z)$ can be written as follows:
 {\allowdisplaybreaks
\begin{align}
&0=\left(3+\z \partial_{\z} \right) \left( \partial^5_z + \frac{20}{z} \partial^4_z +\frac{120}{z^2} \partial^3_z  +\frac{240}{z^3} \partial^2_z +
\frac{120}{z^4} \partial_z  \right) F(z,\z) \nn\\
&~~~~~~~~ + {21 \mu \z \over (1-z)^3} \Big( \partial_z^2 +\frac{8-5 z }{(1-z) z}  \partial_z + \frac{24 z^2-84 z+84 }{7 (1-z)^2 z^2} \Big) F(z,\z) \ .
\end{align}}The equation transforms covariantly under the crossing relation. 

Adopting the function $Q(z,\z)$ and its equation, in this case we have
{\allowdisplaybreaks
\begin{align}
\label{Rm3d}
 Q(z,\z)&= \sum_{n=0} R_n(z) \z^{n+3} \ , ~~~ R_0(z)= z^4  \ , \\
\label{Intm3}
R_{n+1}(z)&=  {-1\over n+4}  \int^{z}_0 ds_5\int^{s_5}_0 ds_4 \int^{s_4}_0 ds_3 \int^{s_3}_0 ds_2 \int^{s_2}_0 ds_1  ~ \nn\\
& ~~~~~~~~~~~~~~~~~~~~~~\times \Bigg[ \frac{ 72 R_n(s_1)}{(1-s_1)^5} + \frac{ 63 R'_n(s_1)}{(1-s_1)^4}  + \frac{ 21 R''_n(s_1)}{(1-s_1)^3}\Bigg] 
\end{align}}where, as before,  we rescale $\z$ to absorb the parameter $\mu$.
Using $R_n = z^4 G_n$, \eqref{Intm3} yields the results consistent with the universal $G_1$, as well as the known 
expressions for $G_2, G_3$ \cite{Kulaxizi:2019tkd, Karlsson:2019dbd, Karlsson:2020ghx}. After computing the higher-order terms of $G_n$ (whose expressions will not be listed here), we again observe patterns in a small $z$ approximation.  At the first three orders, the corresponding $F(z,\z)$ resummation results are given by 
{\allowdisplaybreaks\small \begin{align}
\label{planar3}
&\sum_{n=0}^{\infty} 
\frac{32 \pi   3^{-2 n-\frac{5}{2}}   7^n  \left( -z^3 \z \right)^n}{\Gamma \left(n+\frac{5}{3}\right) \Gamma (n+2) \Gamma \left(n+\frac{7}{3}\right) \Gamma (n+4)} 
= \frac{24}{7 z^3 \z} \Bigg(1- \, _0F_3\left(\Big\{\frac{2}{3},\frac{4}{3},3\Big\} ,\frac{-7 z^3 \z}{9} \right)\Bigg) \ , \\
&\sum_{n=1}^{\infty}  \frac{16 \pi  3^{-2 n-\frac{3}{2}} 7^n n ~z \left(-z^3 \z \right)^n}{\Gamma \left(n+\frac{5}{3}\right) \Gamma (n+2) \Gamma \left(n+\frac{7}{3}\right) \Gamma (n+4)}\nn\\
&=-\frac{ 36}{7 z^2 \z } \Bigg[ 1-2 \, _0F_3\left(\Big\{\frac{2}{3},\frac{4}{3},3\Big\}, \frac{-7 z^3 \z }{9} \right)
+\, _1F_4\left(\Big\{2\Big\}, \Big\{\frac{2}{3},1,\frac{4}{3},3 \Big\}, \frac{-7 z^3 \z}{9} \right) \Bigg] \ , \\
& \sum_{n=1}^{\infty}  \frac{2\ 3^{n+4} 7^{n-1} n\big(7 n (9 n^2+38n +47)+110\big) z^2 \left(-z^3 \z \right)^n}{\Gamma (n+4) \Gamma (3 n+7)} \nn\\
&= -\frac{18 z^2}{5}  \, _0F_3\left(\Big\{\frac{7}{3},\frac{8}{3},3\Big\}, \frac{-7 z^3 \z}{9} \right) -\frac{ 27 z^5 \z}{320} \, _0F_3\left(\Big\{\frac{10}{3},\frac{11}{3},4 \Big\}, \frac{-7 z^3 \z}{9} \right) \nn\\
&~~~~ - \frac{36}{343 z^4 \z^2}  \Bigg[288 \, _0F_3\left(\Big\{\frac{1}{3},\frac{2}{3},1\Big\}, \frac{-7 z^3 \z}{9}\right)
+539 z^3 \z \, _0F_3\left(\Big\{\frac{4}{3},\frac{5}{3},2\Big\}, \frac{-7 z^3 \z}{9} \right)\nn\\
&~~~~~~~~~~~~~~~~~ - 276 \, _0F_3\left(\Big\{\frac{1}{3},\frac{2}{3},2\Big\}, \frac{-7 z^3 \z}{9} \right)-14  z^3 \z-12 \Bigg] \ .
\end{align}}The leading-order result \eqref{planar3} is in agreement with the holographic computation \cite{Fitzpatrick:2019zqz} with  a planar black hole.  The other two expressions are predictions. 

More generally, examining other limits of $z$ also reveal patterns that allow the resummation. 
For instance, by considering $z\to 1$ instead of $z\to 0$, it is also possible to perform the resummation for all $\Delta=-1,-2,-3$ cases. However,  here we refrain from listing excessive expressions based on varying approximations of $z$.

 Without making any approximations of $z$, upon computing the higher-order terms we find that the $d=4$ differential equations suggest a general structure:
 ($X\equiv 1-z$)
{\allowdisplaybreaks\begin{align}
 R_n(z) &= \sum_{j=0,1,2,\dots}^{n} \Big(a_{nj}(\Delta)+  b_{nj}(\Delta) X+  c_{nj}(\Delta) X^2 + d_{nj}(\Delta) X^3  + e_{nj}(\Delta) X^4 \Big) \big(\ln X\big)^j  \nn
\end{align}}where $\Delta=-1,-2,-3$ and the coefficients $a_{nj}, b_{nj}, \dots, e_{nj}$ can be computed to higher orders efficiently.  Integers $n, j$ satisfy $n \geq j \geq 0$. The power of $X$ is restricted. Moreover, we find
\begin{align}
d_{nj}(\Delta=-1) = 0 \ , ~~~~~  e_{nj}(\Delta=-1,-2) =0 \ .
\end{align} 
Note these coefficients are subjected to the initial conditions $\eqref{Rm1d}, \eqref{Rm2d}, \eqref{Rm3d}$. 
We have not yet fully analyzed these coefficients to see if there are any interrelated patterns that would allow a resummation and obtain the closed-form expression without needing to approximate $z$, $i.e.$, a result that may be viewed as a closed-form generalization to \eqref{2dresult}.    
Obtaining such an expansion would be interesting.   As mentioned, information about the Virasoro blocks of general conformal dimensions can be obtained by analytically continuing the correlators involving degenerate operators.
It may be possible to generalize the story to $d=4$ by utilizing the differential equations we consider here.   

\subsection*{4. Discussion}  

In this paper, we have proposed a system of $d=4$ equations \eqref{eqm1}$-$\eqref{eqm3}, valid in the  near-lightcone regime, and verified that the solutions to these equations are in agreement with the existing results obtained via holography and the lightcone bootstrap.  The physical observable we are interested in is the heavy-light scalar four-point correlator at large central charge.   Our approach is guided by the similarity of the $d=2$ BPZ equation in the form of  \eqref{2dBPZ} and crossing symmetry.  It is worth emphasizing that this work is made possible by the advancements in holography and bootstrap.   Let us conclude with additional remarks and speculate on what might lie ahead.

What is the origin of these $d=4$ differential equations?  Could there be an approximate symmetry in the near-lightcone regime? No attempt is made in this work to provide a first-principle derivation, whether through an algebraic approach or otherwise, of the  equations that we observe. 
We do not presume that a rigorous CFT derivation starting with $TT$ and $T{\cal O}$ OPEs can be easily provided. Rather, we suspect that new insights and methods may be needed to fill this logical chasm. Similar to the $d=2$ case, the Ward identities should play a role, but additional ingredients are likely required to rigorously derive these $d=4$ equations. 

There is a clue that arises from a tension. The structure of $d=2$ multi-stress tensors is directly connected to the stress-tensor commutators (which leads to the Virasoro algebra after adopting the Virasoro mode operator).  
In higher dimensions, however, it is known that the $c$-number term (the Schwinger term) of the stress-tensor commutators is divergent, $e.g.$, \cite{Casini:2017roe, Huang:2019fog}. Reconciling this tension may be essential to develop a possible algebraic derivation of these $d=4$ differential equations.\footnote{It has been established that the commutator of the lightray operator in $d=4$ has a divergent central term \cite{Casini:2017roe, Kravchuk:2018htv, Cordova:2018ygx, Belin:2020lsr, Gonzo:2020xza, Besken:2020snx, Korchemsky:2021htm}. A regularization procedure was proposed in \cite{Huang:2021hye} which isolates an algebra-like structure using the integrals over directions transverse to the lightcone (see also \cite{Huang:2019fog, Huang:2020ycs, Huang:2022vcs}). One challenge in utilizing such a formulation is that the effective algebra is made of line operators integrated along the lightcone and their action becomes nonlocal. It is unclear what the right rules for the nonlocal terms should be.  I thank L. Fitzpatrick for this remark.} 

It was discovered that certain $d>2$ superconformal field theories exhibit a chiral algebra\cite{Beem:2013sza, Beem:2014kka, Beem:2014rza}. (See also \cite{Johansen:1994ud, Kapustin:2006hi} which consider extended operators intersecting along a codimension-one surface.) In this work, we focus on the multi-stress tensor sector of the near-lightcone correlator at large $C_T$ without explicitly assuming supersymmetry.  It could be interesting to investigate whether supersymmetry is necessary for these linear differential equations to hold.

Alternatively, it might be possible to derive the null-state equations without prior knowledge of a symmetry. In $d=2$, if one already knows about the fusion rule, one may still derive the level-2 BPZ equation, which corresponds to a rigid Fuchsian system, without knowing the Virasoro algebra. (Such a derivation, however, cannot be extended to the higher-order BPZ equations. See, $e.g.$, \cite{Belavin:2017lvd, Ribault:2014hia} for related remarks.) Another possible derivation of the null-state equations without relying on an algebra is to properly extract the degenerate-operator contributions utilizing only $T\mathcal{O}$ OPE and the Ward identities.  Such an indirect derivation in higher dimensions could be thorny, but valuable.

Since the most significant obstacles arise when transitioning from $d=2$ to any higher dimensions, if the computation can be uplifted from two dimensions to four dimensions, the similar extensions to other $d>2$ dimensions should pose no difficulty. 
For instance, again by drawing upon the $d=2$ BPZ equation in the form of \eqref{2dBPZ}, crossing symmetry, and the input from the universal 
single-stress tensor exchange, we find that in $d=6$, the corresponding equations are  
{{\allowdisplaybreaks\small\begin{align}
&Q_{-1}^{(4,2)}+ 9 { Q_{-1} \over (1-z)^4}=0 \ , \\
&Q_{-2}^{(5,2)}+ 72 {Q_{-2}^{(1,0)} \over (1-z)^4} +144 { Q_{-2} \over (1-z)^5}=0 \ , \\
&Q_{-3}^{(6,2)}+ 324 {Q_{-3}^{(2,0)} \over (1-z)^4} +1296 { Q_{-3}^{(1,0)} \over (1-z)^5}+1800 { Q_{-3} \over (1-z)^6}=0 \ , \\
&Q_{-4}^{(7,2)}+ 1080 {Q_{-4}^{(3,0)} \over (1-z)^4} + 6480 { Q_{-4}^{(2,0)} \over (1-z)^5}+ 18000 { Q_{-4}^{(1,0)} \over (1-z)^6}+21600 { Q_{-4} \over (1-z)^7}=0
\end{align}}}where $Q_{\Delta^*}= {z^2\over (z \z)^{\Delta}} F|_{\Delta= \Delta^*}$ with the special values of $\Delta^*$ in $d=6$ given by $\Delta^*=-1,-2,-3,-4$. We adopt the same convention where the parameter $\mu$ is properly absorbed into the coordinate $\z$ in the near-lightcone expansion. By performing an identical analysis as for the $d=4$ case, we have verified that these equations are consistent with the existing results, obtained via holography and the lightcone bootstrap, for the leading-twist multi-stress tensor contributions. 

However, we would like to mention that the structure of heavy-light correlators in odd dimensions is underinvestigated.  One may observe that there is an increase in the power of the $z$-derivative by one and two when comparing the level-2 BPZ equation with the simplest equations in $d=4$ and $d=6$, respectively, for the $\Delta=-1$ case.  This perhaps suggests that techniques from fractional calculus may assist in organizing the equations in odd dimensions.

Further clarification is needed to establish the validity of the results in this work. These results apply to a class of large-N CFTs that have a holographic dual description as a minimally coupled theory in AdS with arbitrary higher-curvature  corrections, including Einstein gravity as the simplest case. What about non-minimally coupled bulk interactions? 
Here we remark that the corrections to the leading-twist multi-stress tensor OPEs due to the quartic derivative bulk interactions computed in \cite{Fitzpatrick:2020yjb} 
vanish at $\Delta=-1,-2, -3$, which coincide with the required conformal dimensions in the $d=4$ linear differential equations. (To be precise, we are referring to the corrections computed in Eq. (71) and Eq. (86-87) in that paper.) 
A possible future investigation is to search for a mechanism that could explain this phenomenon. 

\subsection*{Acknowledgements}

I am grateful to N. Benjamin, H. Chen, and L. Fitzpatrick for related discussions during the early stage of the work,  to A. Parnachev for helpful remarks on near-lightcone correlators, to F. Haehl and C. Herzog for stimulating conversations. 
 I thank T. Hartman, J. Mann, S. Ribault, K. Skenderis, B. Withers, and M. Yamazaki for interesting comments and discussions.
I also thank Oxford University for their hospitality during the course of this work. 
This research was supported in part by the UKRI grant EP/X030334/1. 

\bibliographystyle{utphys}  
\bibliography{nulleq}  

\providecommand{\href}[2]{#2}\begingroup\raggedright\begin{thebibliography}{100}

\bibitem{BELAVIN1984333}
A.~Belavin, A.~Polyakov, and A.~Zamolodchikov, ``Infinite conformal symmetry in
  two-dimensional quantum field theory,''
  \href{http://dx.doi.org/https://doi.org/10.1016/0550-3213(84)90052-X}{{\em
  Nuclear Physics B} {\bfseries 241} no.~2, (1984) 333--380}.

\bibitem{Fitzpatrick:2012yx}
A.~L. Fitzpatrick, J.~Kaplan, D.~Poland, and D.~Simmons-Duffin, ``{The Analytic
  Bootstrap and AdS Superhorizon Locality},''
  \href{http://dx.doi.org/10.1007/JHEP12(2013)004}{{\em JHEP} {\bfseries 12}
  (2013) 004}, \href{http://arxiv.org/abs/1212.3616}{{\ttfamily arXiv:1212.3616
  [hep-th]}}.

\bibitem{Komargodski:2012ek}
Z.~Komargodski and A.~Zhiboedov, ``{Convexity and Liberation at Large Spin},''
  \href{http://dx.doi.org/10.1007/JHEP11(2013)140}{{\em JHEP} {\bfseries 11}
  (2013) 140}, \href{http://arxiv.org/abs/1212.4103}{{\ttfamily arXiv:1212.4103
  [hep-th]}}.

\bibitem{Alday:2007mf}
L.~F. Alday and J.~M. Maldacena, ``{Comments on operators with large spin},''
  \href{http://dx.doi.org/10.1088/1126-6708/2007/11/019}{{\em JHEP} {\bfseries
  11} (2007) 019}, \href{http://arxiv.org/abs/0708.0672}{{\ttfamily
  arXiv:0708.0672 [hep-th]}}.

\bibitem{Pal:2022vqc}
S.~Pal, J.~Qiao, and S.~Rychkov, ``{Twist accumulation in conformal field
  theory. A rigorous approach to the lightcone bootstrap},''
  \href{http://arxiv.org/abs/2212.04893}{{\ttfamily arXiv:2212.04893
  [hep-th]}}.

\bibitem{zamolodchikov1984conformal}
A.~B. Zamolodchikov, ``Conformal symmetry in two dimensions: an explicit
  recurrence formula for the conformal partial wave amplitude,'' {\em
  Communications in mathematical physics} {\bfseries 96} (1984) 419--422.

\bibitem{zamolodchikov2}
A.~B. Zamolodchikov, ``Conformal symmetry in two-dimensional space: Recursion
  representation of conformal block,'' {\em Theor Math Phys} {\bfseries 73}
  (1987) 1088–1093.

\bibitem{Harlow:2011ny}
D.~Harlow, J.~Maltz, and E.~Witten, ``{Analytic Continuation of Liouville
  Theory},'' \href{http://dx.doi.org/10.1007/JHEP12(2011)071}{{\em JHEP}
  {\bfseries 12} (2011) 071}, \href{http://arxiv.org/abs/1108.4417}{{\ttfamily
  arXiv:1108.4417 [hep-th]}}.

\bibitem{Hartman:2013mia}
T.~Hartman, ``{Entanglement Entropy at Large Central Charge},''
  \href{http://arxiv.org/abs/1303.6955}{{\ttfamily arXiv:1303.6955 [hep-th]}}.

\bibitem{Litvinov:2013sxa}
A.~Litvinov, S.~Lukyanov, N.~Nekrasov, and A.~Zamolodchikov, ``{Classical
  Conformal Blocks and Painleve VI},''
  \href{http://dx.doi.org/10.1007/JHEP07(2014)144}{{\em JHEP} {\bfseries 07}
  (2014) 144}, \href{http://arxiv.org/abs/1309.4700}{{\ttfamily arXiv:1309.4700
  [hep-th]}}.

\bibitem{Fitzpatrick:2014vua}
A.~L. Fitzpatrick, J.~Kaplan, and M.~T. Walters, ``{Universality of
  Long-Distance AdS Physics from the CFT Bootstrap},''
  \href{http://dx.doi.org/10.1007/JHEP08(2014)145}{{\em JHEP} {\bfseries 08}
  (2014) 145}, \href{http://arxiv.org/abs/1403.6829}{{\ttfamily arXiv:1403.6829
  [hep-th]}}.

\bibitem{Asplund:2014coa}
C.~T. Asplund, A.~Bernamonti, F.~Galli, and T.~Hartman, ``{Holographic
  Entanglement Entropy from 2d CFT: Heavy States and Local Quenches},''
  \href{http://dx.doi.org/10.1007/JHEP02(2015)171}{{\em JHEP} {\bfseries 02}
  (2015) 171}, \href{http://arxiv.org/abs/1410.1392}{{\ttfamily arXiv:1410.1392
  [hep-th]}}.

\bibitem{Roberts:2014ifa}
D.~A. Roberts and D.~Stanford, ``{Two-dimensional conformal field theory and
  the butterfly effect},''
  \href{http://dx.doi.org/10.1103/PhysRevLett.115.131603}{{\em Phys. Rev.
  Lett.} {\bfseries 115} no.~13, (2015) 131603},
  \href{http://arxiv.org/abs/1412.5123}{{\ttfamily arXiv:1412.5123 [hep-th]}}.

\bibitem{Hijano:2015rla}
E.~Hijano, P.~Kraus, and R.~Snively, ``{Worldline approach to semi-classical
  conformal blocks},'' \href{http://dx.doi.org/10.1007/JHEP07(2015)131}{{\em
  JHEP} {\bfseries 07} (2015) 131},
  \href{http://arxiv.org/abs/1501.02260}{{\ttfamily arXiv:1501.02260
  [hep-th]}}.

\bibitem{Fitzpatrick:2015zha}
A.~L. Fitzpatrick, J.~Kaplan, and M.~T. Walters, ``{Virasoro Conformal Blocks
  and Thermality from Classical Background Fields},''
  \href{http://dx.doi.org/10.1007/JHEP11(2015)200}{{\em JHEP} {\bfseries 11}
  (2015) 200}, \href{http://arxiv.org/abs/1501.05315}{{\ttfamily
  arXiv:1501.05315 [hep-th]}}.

\bibitem{Perlmutter:2015iya}
E.~Perlmutter, ``{Virasoro conformal blocks in closed form},''
  \href{http://dx.doi.org/10.1007/JHEP08(2015)088}{{\em JHEP} {\bfseries 08}
  (2015) 088}, \href{http://arxiv.org/abs/1502.07742}{{\ttfamily
  arXiv:1502.07742 [hep-th]}}.

\bibitem{Hijano:2015qja}
E.~Hijano, P.~Kraus, E.~Perlmutter, and R.~Snively, ``{Semiclassical Virasoro
  blocks from AdS$_{3}$ gravity},''
  \href{http://dx.doi.org/10.1007/JHEP12(2015)077}{{\em JHEP} {\bfseries 12}
  (2015) 077}, \href{http://arxiv.org/abs/1508.04987}{{\ttfamily
  arXiv:1508.04987 [hep-th]}}.

\bibitem{Fitzpatrick:2015foa}
A.~L. Fitzpatrick, J.~Kaplan, M.~T. Walters, and J.~Wang, ``{Hawking from
  Catalan},'' \href{http://dx.doi.org/10.1007/JHEP05(2016)069}{{\em JHEP}
  {\bfseries 05} (2016) 069}, \href{http://arxiv.org/abs/1510.00014}{{\ttfamily
  arXiv:1510.00014 [hep-th]}}.

\bibitem{Beccaria:2015shq}
M.~Beccaria, A.~Fachechi, and G.~Macorini, ``{Virasoro vacuum block at
  next-to-leading order in the heavy-light limit},''
  \href{http://dx.doi.org/10.1007/JHEP02(2016)072}{{\em JHEP} {\bfseries 02}
  (2016) 072}, \href{http://arxiv.org/abs/1511.05452}{{\ttfamily
  arXiv:1511.05452 [hep-th]}}.

\bibitem{Fitzpatrick:2015dlt}
A.~L. Fitzpatrick and J.~Kaplan, ``{Conformal Blocks Beyond the Semi-Classical
  Limit},'' \href{http://dx.doi.org/10.1007/JHEP05(2016)075}{{\em JHEP}
  {\bfseries 05} (2016) 075}, \href{http://arxiv.org/abs/1512.03052}{{\ttfamily
  arXiv:1512.03052 [hep-th]}}.

\bibitem{Alkalaev:2015fbw}
K.~B. Alkalaev and V.~A. Belavin, ``{From global to heavy-light: 5-point
  conformal blocks},'' \href{http://dx.doi.org/10.1007/JHEP03(2016)184}{{\em
  JHEP} {\bfseries 03} (2016) 184},
  \href{http://arxiv.org/abs/1512.07627}{{\ttfamily arXiv:1512.07627
  [hep-th]}}.

\bibitem{Fitzpatrick:2016thx}
A.~L. Fitzpatrick and J.~Kaplan, ``{A Quantum Correction To Chaos},''
  \href{http://dx.doi.org/10.1007/JHEP05(2016)070}{{\em JHEP} {\bfseries 05}
  (2016) 070}, \href{http://arxiv.org/abs/1601.06164}{{\ttfamily
  arXiv:1601.06164 [hep-th]}}.

\bibitem{Banerjee:2016qca}
P.~Banerjee, S.~Datta, and R.~Sinha, ``{Higher-point conformal blocks and
  entanglement entropy in heavy states},''
  \href{http://dx.doi.org/10.1007/JHEP05(2016)127}{{\em JHEP} {\bfseries 05}
  (2016) 127}, \href{http://arxiv.org/abs/1601.06794}{{\ttfamily
  arXiv:1601.06794 [hep-th]}}.

\bibitem{Anous:2016kss}
T.~Anous, T.~Hartman, A.~Rovai, and J.~Sonner, ``{Black Hole Collapse in the
  1/c Expansion},'' \href{http://dx.doi.org/10.1007/JHEP07(2016)123}{{\em JHEP}
  {\bfseries 07} (2016) 123}, \href{http://arxiv.org/abs/1603.04856}{{\ttfamily
  arXiv:1603.04856 [hep-th]}}.

\bibitem{Fitzpatrick:2016ive}
A.~L. Fitzpatrick, J.~Kaplan, D.~Li, and J.~Wang, ``{On information loss in
  AdS$_{3}$/CFT$_{2}$},'' \href{http://dx.doi.org/10.1007/JHEP05(2016)109}{{\em
  JHEP} {\bfseries 05} (2016) 109},
  \href{http://arxiv.org/abs/1603.08925}{{\ttfamily arXiv:1603.08925
  [hep-th]}}.

\bibitem{Chen:2016cms}
H.~Chen, A.~L. Fitzpatrick, J.~Kaplan, D.~Li, and J.~Wang, ``{Degenerate
  Operators and the $1/c$ Expansion: Lorentzian Resummations, High Order
  Computations, and Super-Virasoro Blocks},''
  \href{http://dx.doi.org/10.1007/JHEP03(2017)167}{{\em JHEP} {\bfseries 03}
  (2017) 167}, \href{http://arxiv.org/abs/1606.02659}{{\ttfamily
  arXiv:1606.02659 [hep-th]}}.

\bibitem{Chen:2016dfb}
B.~Chen, J.-q. Wu, and J.-j. Zhang, ``{Holographic Description of 2D Conformal
  Block in Semi-classical Limit},''
  \href{http://dx.doi.org/10.1007/JHEP10(2016)110}{{\em JHEP} {\bfseries 10}
  (2016) 110}, \href{http://arxiv.org/abs/1609.00801}{{\ttfamily
  arXiv:1609.00801 [hep-th]}}.

\bibitem{Maloney:2016kee}
A.~Maloney, H.~Maxfield, and G.~S. Ng, ``{A conformal block Farey tail},''
  \href{http://dx.doi.org/10.1007/JHEP06(2017)117}{{\em JHEP} {\bfseries 06}
  (2017) 117}, \href{http://arxiv.org/abs/1609.02165}{{\ttfamily
  arXiv:1609.02165 [hep-th]}}.

\bibitem{Fitzpatrick:2016mtp}
A.~L. Fitzpatrick, J.~Kaplan, D.~Li, and J.~Wang, ``{Exact Virasoro Blocks from
  Wilson Lines and Background-Independent Operators},''
  \href{http://dx.doi.org/10.1007/JHEP07(2017)092}{{\em JHEP} {\bfseries 07}
  (2017) 092}, \href{http://arxiv.org/abs/1612.06385}{{\ttfamily
  arXiv:1612.06385 [hep-th]}}.

\bibitem{Chen:2017yze}
H.~Chen, C.~Hussong, J.~Kaplan, and D.~Li, ``{A Numerical Approach to Virasoro
  Blocks and the Information Paradox},''
  \href{http://dx.doi.org/10.1007/JHEP09(2017)102}{{\em JHEP} {\bfseries 09}
  (2017) 102}, \href{http://arxiv.org/abs/1703.09727}{{\ttfamily
  arXiv:1703.09727 [hep-th]}}.

\bibitem{Cho:2017oxl}
M.~Cho, S.~Collier, and X.~Yin, ``{Recursive Representations of Arbitrary
  Virasoro Conformal Blocks},''
  \href{http://dx.doi.org/10.1007/JHEP04(2019)018}{{\em JHEP} {\bfseries 04}
  (2019) 018}, \href{http://arxiv.org/abs/1703.09805}{{\ttfamily
  arXiv:1703.09805 [hep-th]}}.

\bibitem{Cho:2017fzo}
M.~Cho, S.~Collier, and X.~Yin, ``{Genus Two Modular Bootstrap},''
  \href{http://dx.doi.org/10.1007/JHEP04(2019)022}{{\em JHEP} {\bfseries 04}
  (2019) 022}, \href{http://arxiv.org/abs/1705.05865}{{\ttfamily
  arXiv:1705.05865 [hep-th]}}.

\bibitem{Galliani:2017jlg}
A.~Galliani, S.~Giusto, and R.~Russo, ``{Holographic 4-point correlators with
  heavy states},'' \href{http://dx.doi.org/10.1007/JHEP10(2017)040}{{\em JHEP}
  {\bfseries 10} (2017) 040}, \href{http://arxiv.org/abs/1705.09250}{{\ttfamily
  arXiv:1705.09250 [hep-th]}}.

\bibitem{Bombini:2017sge}
A.~Bombini, A.~Galliani, S.~Giusto, E.~Moscato, and R.~Russo, ``{Unitary
  4-point correlators from classical geometries},''
  \href{http://dx.doi.org/10.1140/epjc/s10052-017-5492-3}{{\em Eur. Phys. J. C}
  {\bfseries 78} no.~1, (2018) 8},
  \href{http://arxiv.org/abs/1710.06820}{{\ttfamily arXiv:1710.06820
  [hep-th]}}.

\bibitem{Kusuki:2018nms}
Y.~Kusuki, ``{Large $c$ Virasoro Blocks from Monodromy Method beyond Known
  Limits},'' \href{http://dx.doi.org/10.1007/JHEP08(2018)161}{{\em JHEP}
  {\bfseries 08} (2018) 161}, \href{http://arxiv.org/abs/1806.04352}{{\ttfamily
  arXiv:1806.04352 [hep-th]}}.

\bibitem{Bombini:2018jrg}
A.~Bombini, S.~Giusto, and R.~Russo, ``{A note on the Virasoro blocks at order
  $1/c$},'' \href{http://dx.doi.org/10.1140/epjc/s10052-018-6522-5}{{\em Eur.
  Phys. J. C} {\bfseries 79} no.~1, (2019) 3},
  \href{http://arxiv.org/abs/1807.07886}{{\ttfamily arXiv:1807.07886
  [hep-th]}}.

\bibitem{Cotler:2018zff}
J.~Cotler and K.~Jensen, ``{A theory of reparameterizations for AdS$_3$
  gravity},'' \href{http://dx.doi.org/10.1007/JHEP02(2019)079}{{\em JHEP}
  {\bfseries 02} (2019) 079}, \href{http://arxiv.org/abs/1808.03263}{{\ttfamily
  arXiv:1808.03263 [hep-th]}}.

\bibitem{Kusuki:2018wpa}
Y.~Kusuki, ``{Light Cone Bootstrap in General 2D CFTs and Entanglement from
  Light Cone Singularity},''
  \href{http://dx.doi.org/10.1007/JHEP01(2019)025}{{\em JHEP} {\bfseries 01}
  (2019) 025}, \href{http://arxiv.org/abs/1810.01335}{{\ttfamily
  arXiv:1810.01335 [hep-th]}}.

\bibitem{Collier:2018exn}
S.~Collier, Y.~Gobeil, H.~Maxfield, and E.~Perlmutter, ``{Quantum Regge
  Trajectories and the Virasoro Analytic Bootstrap},''
  \href{http://dx.doi.org/10.1007/JHEP05(2019)212}{{\em JHEP} {\bfseries 05}
  (2019) 212}, \href{http://arxiv.org/abs/1811.05710}{{\ttfamily
  arXiv:1811.05710 [hep-th]}}.

\bibitem{Kulaxizi:2018dxo}
M.~Kulaxizi, G.~S. Ng, and A.~Parnachev, ``{Black Holes, Heavy States, Phase
  Shift and Anomalous Dimensions},''
  \href{http://dx.doi.org/10.21468/SciPostPhys.6.6.065}{{\em SciPost Phys.}
  {\bfseries 6} no.~6, (2019) 065},
  \href{http://arxiv.org/abs/1812.03120}{{\ttfamily arXiv:1812.03120
  [hep-th]}}.

\bibitem{Chang:2018nzm}
C.-M. Chang, D.~M. Ramirez, and M.~Rangamani, ``{Spinning constraints on
  chaotic large $c$ CFTs},''
  \href{http://dx.doi.org/10.1007/JHEP03(2019)068}{{\em JHEP} {\bfseries 03}
  (2019) 068}, \href{http://arxiv.org/abs/1812.05585}{{\ttfamily
  arXiv:1812.05585 [hep-th]}}.

\bibitem{Giusto:2018ovt}
S.~Giusto, R.~Russo, and C.~Wen, ``{Holographic correlators in AdS$_{3}$},''
  \href{http://dx.doi.org/10.1007/JHEP03(2019)096}{{\em JHEP} {\bfseries 03}
  (2019) 096}, \href{http://arxiv.org/abs/1812.06479}{{\ttfamily
  arXiv:1812.06479 [hep-th]}}.

\bibitem{Anous:2019yku}
T.~Anous and J.~Sonner, ``{Phases of scrambling in eigenstates},''
  \href{http://dx.doi.org/10.21468/SciPostPhys.7.1.003}{{\em SciPost Phys.}
  {\bfseries 7} (2019) 003}, \href{http://arxiv.org/abs/1903.03143}{{\ttfamily
  arXiv:1903.03143 [hep-th]}}.

\bibitem{Kusuki:2019gjs}
Y.~Kusuki and M.~Miyaji, ``{Entanglement Entropy, OTOC and Bootstrap in 2D CFTs
  from Regge and Light Cone Limits of Multi-point Conformal Block},''
  \href{http://dx.doi.org/10.1007/JHEP08(2019)063}{{\em JHEP} {\bfseries 08}
  (2019) 063}, \href{http://arxiv.org/abs/1905.02191}{{\ttfamily
  arXiv:1905.02191 [hep-th]}}.

\bibitem{Besken:2019bsu}
M.~Be\c{s}ken, S.~Datta, and P.~Kraus, ``{Quantum thermalization and Virasoro
  symmetry},'' \href{http://dx.doi.org/10.1088/1742-5468/ab900b}{{\em J. Stat.
  Mech.} {\bfseries 2006} (2020) 063104},
  \href{http://arxiv.org/abs/1907.06661}{{\ttfamily arXiv:1907.06661
  [hep-th]}}.

\bibitem{Haehl:2019eae}
F.~M. Haehl, W.~Reeves, and M.~Rozali, ``{Reparametrization modes, shadow
  operators, and quantum chaos in higher-dimensional CFTs},''
  \href{http://dx.doi.org/10.1007/JHEP11(2019)102}{{\em JHEP} {\bfseries 11}
  (2019) 102}, \href{http://arxiv.org/abs/1909.05847}{{\ttfamily
  arXiv:1909.05847 [hep-th]}}.

\bibitem{Besken:2019jyw}
M.~Be\c{s}ken, S.~Datta, and P.~Kraus, ``{Semi-classical Virasoro blocks: proof
  of exponentiation},'' \href{http://dx.doi.org/10.1007/JHEP01(2020)109}{{\em
  JHEP} {\bfseries 01} (2020) 109},
  \href{http://arxiv.org/abs/1910.04169}{{\ttfamily arXiv:1910.04169
  [hep-th]}}.

\bibitem{Collier:2019weq}
S.~Collier, A.~Maloney, H.~Maxfield, and I.~Tsiares, ``{Universal dynamics of
  heavy operators in CFT$_{2}$},''
  \href{http://dx.doi.org/10.1007/JHEP07(2020)074}{{\em JHEP} {\bfseries 07}
  (2020) 074}, \href{http://arxiv.org/abs/1912.00222}{{\ttfamily
  arXiv:1912.00222 [hep-th]}}.

\bibitem{Alkalaev:2020kxz}
K.~Alkalaev and M.~Pavlov, ``{Holographic variables for CFT$_2$ conformal
  blocks with heavy operators},''
  \href{http://dx.doi.org/10.1016/j.nuclphysb.2020.115018}{{\em Nucl. Phys. B}
  {\bfseries 956} (2020) 115018},
  \href{http://arxiv.org/abs/2001.02604}{{\ttfamily arXiv:2001.02604
  [hep-th]}}.

\bibitem{Anous:2020vtw}
T.~Anous and F.~M. Haehl, ``{On the Virasoro six-point identity block and
  chaos},'' \href{http://dx.doi.org/10.1007/JHEP08(2020)002}{{\em JHEP}
  {\bfseries 08} no.~08, (2020) 002},
  \href{http://arxiv.org/abs/2005.06440}{{\ttfamily arXiv:2005.06440
  [hep-th]}}.

\bibitem{Das:2020fhs}
D.~Das, S.~Datta, and M.~Raman, ``{Virasoro blocks and quasimodular forms},''
  \href{http://dx.doi.org/10.1007/JHEP11(2020)010}{{\em JHEP} {\bfseries 11}
  (2020) 010}, \href{http://arxiv.org/abs/2007.10998}{{\ttfamily
  arXiv:2007.10998 [hep-th]}}.

\bibitem{Nguyen:2021jja}
K.~Nguyen, ``{Reparametrization modes in 2d CFT and the effective theory of
  stress tensor exchanges},''
  \href{http://dx.doi.org/10.1007/JHEP05(2021)029}{{\em JHEP} {\bfseries 05}
  (2021) 029}, \href{http://arxiv.org/abs/2101.08800}{{\ttfamily
  arXiv:2101.08800 [hep-th]}}.

\bibitem{Nguyen:2022xsw}
K.~Nguyen, ``{Virasoro blocks and the reparametrization formalism},''
  \href{http://dx.doi.org/10.1007/JHEP04(2023)143}{{\em JHEP} {\bfseries 04}
  (2023) 143}, \href{http://arxiv.org/abs/2212.02527}{{\ttfamily
  arXiv:2212.02527 [hep-th]}}.

\bibitem{Benjamin:2023uib}
N.~Benjamin, S.~Collier, A.~Maloney, and V.~Meruliya, ``{Resurgence, conformal
  blocks, and the sum over geometries in quantum gravity},''
  \href{http://dx.doi.org/10.1007/JHEP05(2023)166}{{\em JHEP} {\bfseries 05}
  (2023) 166}, \href{http://arxiv.org/abs/2302.12851}{{\ttfamily
  arXiv:2302.12851 [hep-th]}}.

\bibitem{Heemskerk:2009pn}
I.~Heemskerk, J.~Penedones, J.~Polchinski, and J.~Sully, ``{Holography from
  Conformal Field Theory},''
  \href{http://dx.doi.org/10.1088/1126-6708/2009/10/079}{{\em JHEP} {\bfseries
  10} (2009) 079}, \href{http://arxiv.org/abs/0907.0151}{{\ttfamily
  arXiv:0907.0151 [hep-th]}}.

\bibitem{Fitzpatrick:2010zm}
A.~L. Fitzpatrick, E.~Katz, D.~Poland, and D.~Simmons-Duffin, ``{Effective
  Conformal Theory and the Flat-Space Limit of AdS},''
  \href{http://dx.doi.org/10.1007/JHEP07(2011)023}{{\em JHEP} {\bfseries 07}
  (2011) 023}, \href{http://arxiv.org/abs/1007.2412}{{\ttfamily arXiv:1007.2412
  [hep-th]}}.

\bibitem{Camanho:2014apa}
X.~O. Camanho, J.~D. Edelstein, J.~Maldacena, and A.~Zhiboedov, ``{Causality
  Constraints on Corrections to the Graviton Three-Point Coupling},''
  \href{http://dx.doi.org/10.1007/JHEP02(2016)020}{{\em JHEP} {\bfseries 02}
  (2016) 020}, \href{http://arxiv.org/abs/1407.5597}{{\ttfamily arXiv:1407.5597
  [hep-th]}}.

\bibitem{Hartman:2015lfa}
T.~Hartman, S.~Jain, and S.~Kundu, ``{Causality Constraints in Conformal Field
  Theory},'' \href{http://dx.doi.org/10.1007/JHEP05(2016)099}{{\em JHEP}
  {\bfseries 05} (2016) 099}, \href{http://arxiv.org/abs/1509.00014}{{\ttfamily
  arXiv:1509.00014 [hep-th]}}.

\bibitem{Komargodski:2016gci}
Z.~Komargodski, M.~Kulaxizi, A.~Parnachev, and A.~Zhiboedov, ``{Conformal Field
  Theories and Deep Inelastic Scattering},''
  \href{http://dx.doi.org/10.1103/PhysRevD.95.065011}{{\em Phys. Rev. D}
  {\bfseries 95} no.~6, (2017) 065011},
  \href{http://arxiv.org/abs/1601.05453}{{\ttfamily arXiv:1601.05453
  [hep-th]}}.

\bibitem{Afkhami-Jeddi:2016ntf}
N.~Afkhami-Jeddi, T.~Hartman, S.~Kundu, and A.~Tajdini, ``{Einstein gravity
  3-point functions from conformal field theory},''
  \href{http://dx.doi.org/10.1007/JHEP12(2017)049}{{\em JHEP} {\bfseries 12}
  (2017) 049}, \href{http://arxiv.org/abs/1610.09378}{{\ttfamily
  arXiv:1610.09378 [hep-th]}}.

\bibitem{Kulaxizi:2017ixa}
M.~Kulaxizi, A.~Parnachev, and A.~Zhiboedov, ``{Bulk Phase Shift, CFT Regge
  Limit and Einstein Gravity},''
  \href{http://dx.doi.org/10.1007/JHEP06(2018)121}{{\em JHEP} {\bfseries 06}
  (2018) 121}, \href{http://arxiv.org/abs/1705.02934}{{\ttfamily
  arXiv:1705.02934 [hep-th]}}.

\bibitem{Li:2017lmh}
D.~Li, D.~Meltzer, and D.~Poland, ``{Conformal Bootstrap in the Regge Limit},''
  \href{http://dx.doi.org/10.1007/JHEP12(2017)013}{{\em JHEP} {\bfseries 12}
  (2017) 013}, \href{http://arxiv.org/abs/1705.03453}{{\ttfamily
  arXiv:1705.03453 [hep-th]}}.

\bibitem{Meltzer:2017rtf}
D.~Meltzer and E.~Perlmutter, ``{Beyond $a = c$: gravitational couplings to
  matter and the stress tensor OPE},''
  \href{http://dx.doi.org/10.1007/JHEP07(2018)157}{{\em JHEP} {\bfseries 07}
  (2018) 157}, \href{http://arxiv.org/abs/1712.04861}{{\ttfamily
  arXiv:1712.04861 [hep-th]}}.

\bibitem{Afkhami-Jeddi:2018apj}
N.~Afkhami-Jeddi, S.~Kundu, and A.~Tajdini, ``{A Bound on Massive Higher Spin
  Particles},'' \href{http://dx.doi.org/10.1007/JHEP04(2019)056}{{\em JHEP}
  {\bfseries 04} (2019) 056}, \href{http://arxiv.org/abs/1811.01952}{{\ttfamily
  arXiv:1811.01952 [hep-th]}}.

\bibitem{Belin:2019mnx}
A.~Belin, D.~M. Hofman, and G.~Mathys, ``{Einstein gravity from ANEC
  correlators},'' \href{http://dx.doi.org/10.1007/JHEP08(2019)032}{{\em JHEP}
  {\bfseries 08} (2019) 032}, \href{http://arxiv.org/abs/1904.05892}{{\ttfamily
  arXiv:1904.05892 [hep-th]}}.

\bibitem{Kologlu:2019bco}
M.~Kologlu, P.~Kravchuk, D.~Simmons-Duffin, and A.~Zhiboedov, ``{Shocks,
  Superconvergence, and a Stringy Equivalence Principle},''
  \href{http://dx.doi.org/10.1007/JHEP11(2020)096}{{\em JHEP} {\bfseries 11}
  (2020) 096}, \href{http://arxiv.org/abs/1904.05905}{{\ttfamily
  arXiv:1904.05905 [hep-th]}}.

\bibitem{Caron-Huot:2021enk}
S.~Caron-Huot, D.~Mazac, L.~Rastelli, and D.~Simmons-Duffin, ``{AdS bulk
  locality from sharp CFT bounds},''
  \href{http://dx.doi.org/10.1007/JHEP11(2021)164}{{\em JHEP} {\bfseries 11}
  (2021) 164}, \href{http://arxiv.org/abs/2106.10274}{{\ttfamily
  arXiv:2106.10274 [hep-th]}}.

\bibitem{Maldacena:1997re}
J.~M. Maldacena, ``{The Large N limit of superconformal field theories and
  supergravity},'' \href{http://dx.doi.org/10.4310/ATMP.1998.v2.n2.a1}{{\em
  Adv. Theor. Math. Phys.} {\bfseries 2} (1998) 231--252},
  \href{http://arxiv.org/abs/hep-th/9711200}{{\ttfamily arXiv:hep-th/9711200}}.

\bibitem{Gubser:1998bc}
S.~S. Gubser, I.~R. Klebanov, and A.~M. Polyakov, ``{Gauge theory correlators
  from noncritical string theory},''
  \href{http://dx.doi.org/10.1016/S0370-2693(98)00377-3}{{\em Phys. Lett. B}
  {\bfseries 428} (1998) 105--114},
  \href{http://arxiv.org/abs/hep-th/9802109}{{\ttfamily arXiv:hep-th/9802109}}.

\bibitem{Witten:1998qj}
E.~Witten, ``{Anti-de Sitter space and holography},''
  \href{http://dx.doi.org/10.4310/ATMP.1998.v2.n2.a2}{{\em Adv. Theor. Math.
  Phys.} {\bfseries 2} (1998) 253--291},
  \href{http://arxiv.org/abs/hep-th/9802150}{{\ttfamily arXiv:hep-th/9802150}}.

\bibitem{Banados:1992wn}
M.~Banados, C.~Teitelboim, and J.~Zanelli, ``{The Black hole in
  three-dimensional space-time},''
  \href{http://dx.doi.org/10.1103/PhysRevLett.69.1849}{{\em Phys. Rev. Lett.}
  {\bfseries 69} (1992) 1849--1851},
  \href{http://arxiv.org/abs/hep-th/9204099}{{\ttfamily arXiv:hep-th/9204099}}.

\bibitem{Keski-Vakkuri:1998gmz}
E.~Keski-Vakkuri, ``{Bulk and boundary dynamics in BTZ black holes},''
  \href{http://dx.doi.org/10.1103/PhysRevD.59.104001}{{\em Phys. Rev. D}
  {\bfseries 59} (1999) 104001},
  \href{http://arxiv.org/abs/hep-th/9808037}{{\ttfamily arXiv:hep-th/9808037}}.

\bibitem{Fitzpatrick:2019zqz}
A.~L. Fitzpatrick and K.-W. Huang, ``{Universal Lowest-Twist in CFTs from
  Holography},'' \href{http://dx.doi.org/10.1007/JHEP08(2019)138}{{\em JHEP}
  {\bfseries 08} (2019) 138},
\href{http://arxiv.org/abs/1903.05306}{{\ttfamily arXiv:1903.05306 [hep-th]}}.

\bibitem{Fitzpatrick:2019efk}
A.~L. Fitzpatrick, K.-W. Huang, and D.~Li, ``{Probing universalities in $d > 2$
  CFTs: from black holes to shockwaves},''
  \href{http://dx.doi.org/10.1007/JHEP11(2019)139}{{\em JHEP} {\bfseries 11}
  (2019) 139},
\href{http://arxiv.org/abs/1907.10810}{{\ttfamily arXiv:1907.10810 [hep-th]}}.

\bibitem{Fitzpatrick:2020yjb}
A.~L. Fitzpatrick, K.-W. Huang, D.~Meltzer, E.~Perlmutter, and
  D.~Simmons-Duffin, ``{Model-dependence of minimal-twist OPEs in d
  \ensuremath{>} 2 holographic CFTs},''
  \href{http://dx.doi.org/10.1007/JHEP11(2020)060}{{\em JHEP} {\bfseries 11}
  (2020) 060}, \href{http://arxiv.org/abs/2007.07382}{{\ttfamily
  arXiv:2007.07382 [hep-th]}}.

\bibitem{Kulaxizi:2019tkd}
M.~Kulaxizi, G.~S. Ng, and A.~Parnachev, ``{Subleading Eikonal, AdS/CFT and
  Double Stress Tensors},''
  \href{http://dx.doi.org/10.1007/JHEP10(2019)107}{{\em JHEP} {\bfseries 10}
  (2019) 107},
\href{http://arxiv.org/abs/1907.00867}{{\ttfamily arXiv:1907.00867 [hep-th]}}.

\bibitem{Karlsson:2019dbd}
R.~Karlsson, M.~Kulaxizi, A.~Parnachev, and P.~Tadić, ``{Leading Multi-Stress
  Tensors and Conformal Bootstrap},''
  \href{http://dx.doi.org/10.1007/JHEP01(2020)076}{{\em JHEP} {\bfseries 01}
  (2020) 076},
\href{http://arxiv.org/abs/1909.05775}{{\ttfamily arXiv:1909.05775 [hep-th]}}.

\bibitem{Karlsson:2020ghx}
R.~Karlsson, M.~Kulaxizi, A.~Parnachev, and P.~Tadi\'c, ``{Stress tensor sector
  of conformal correlators operators in the Regge limit},''
  \href{http://dx.doi.org/10.1007/JHEP07(2020)019}{{\em JHEP} {\bfseries 07}
  (2020) 019}, \href{http://arxiv.org/abs/2002.12254}{{\ttfamily
  arXiv:2002.12254 [hep-th]}}.

\bibitem{Huang:2021hye}
K.-W. Huang, ``{$d>2$ stress-tensor operator product expansion near a line},''
  \href{http://dx.doi.org/10.1103/PhysRevD.103.L121702}{{\em Phys. Rev. D}
  {\bfseries 103} no.~12, (2021) 121702},
  \href{http://arxiv.org/abs/2103.09930}{{\ttfamily arXiv:2103.09930
  [hep-th]}}.

\bibitem{Karlsson:2021mgg}
R.~Karlsson, M.~Kulaxizi, G.~S. Ng, A.~Parnachev, and P.~Tadi\'c, ``{CFT
  correlators, ${\mathcal W}$-algebras and Generalized Catalan Numbers},''
  \href{http://arxiv.org/abs/2111.07924}{{\ttfamily arXiv:2111.07924
  [hep-th]}}.

\bibitem{Caron-Huot:2017vep}
S.~Caron-Huot, ``{Analyticity in Spin in Conformal Theories},''
  \href{http://dx.doi.org/10.1007/JHEP09(2017)078}{{\em JHEP} {\bfseries 09}
  (2017) 078}, \href{http://arxiv.org/abs/1703.00278}{{\ttfamily
  arXiv:1703.00278 [hep-th]}}.

\bibitem{Simmons-Duffin:2017nub}
D.~Simmons-Duffin, D.~Stanford, and E.~Witten, ``{A spacetime derivation of the
  Lorentzian OPE inversion formula},''
  \href{http://dx.doi.org/10.1007/JHEP07(2018)085}{{\em JHEP} {\bfseries 07}
  (2018) 085}, \href{http://arxiv.org/abs/1711.03816}{{\ttfamily
  arXiv:1711.03816 [hep-th]}}.

\bibitem{Li:2019zba}
Y.-Z. Li, ``{Heavy-light Bootstrap from Lorentzian Inversion Formula},''
  \href{http://dx.doi.org/10.1007/JHEP07(2020)046}{{\em JHEP} {\bfseries 07}
  (2020) 046}, \href{http://arxiv.org/abs/1910.06357}{{\ttfamily
  arXiv:1910.06357 [hep-th]}}.

\bibitem{Li:2020dqm}
Y.-Z. Li and H.-Y. Zhang, ``{More on heavy-light bootstrap up to
  double-stress-tensor},''
  \href{http://dx.doi.org/10.1007/JHEP10(2020)055}{{\em JHEP} {\bfseries 10}
  (2020) 055}, \href{http://arxiv.org/abs/2004.04758}{{\ttfamily
  arXiv:2004.04758 [hep-th]}}.

\bibitem{Parnachev:2020fna}
A.~Parnachev, ``{Near Lightcone Thermal Conformal Correlators and
  Holography},'' \href{http://arxiv.org/abs/2005.06877}{{\ttfamily
  arXiv:2005.06877 [hep-th]}}.

\bibitem{Karlsson:2021duj}
R.~Karlsson, A.~Parnachev, and P.~Tadi\'c, ``{Thermalization in large-N
  CFTs},'' \href{http://dx.doi.org/10.1007/JHEP09(2021)205}{{\em JHEP}
  {\bfseries 09} (2021) 205}, \href{http://arxiv.org/abs/2102.04953}{{\ttfamily
  arXiv:2102.04953 [hep-th]}}.

\bibitem{Dodelson:2022yvn}
M.~Dodelson, A.~Grassi, C.~Iossa, D.~Panea~Lichtig, and A.~Zhiboedov,
  ``{Holographic thermal correlators from supersymmetric instantons},''
  \href{http://dx.doi.org/10.21468/SciPostPhys.14.5.116}{{\em SciPost Phys.}
  {\bfseries 14} (2023) 116}, \href{http://arxiv.org/abs/2206.07720}{{\ttfamily
  arXiv:2206.07720 [hep-th]}}.

\bibitem{Nekrasov:2009rc}
N.~A. Nekrasov and S.~L. Shatashvili,
  \href{http://dx.doi.org/10.1142/9789814304634_0015}{``{Quantization of
  Integrable Systems and Four Dimensional Gauge Theories},''} in {\em {16th
  International Congress on Mathematical Physics}}, pp.~265--289.
\newblock 8, 2009.
\newblock \href{http://arxiv.org/abs/0908.4052}{{\ttfamily arXiv:0908.4052
  [hep-th]}}.

\bibitem{Casini:2017roe}
H.~Casini, E.~Teste, and G.~Torroba, ``{Modular Hamiltonians on the null plane
  and the Markov property of the vacuum state},''
  \href{http://dx.doi.org/10.1088/1751-8121/aa7eaa}{{\em J. Phys. A} {\bfseries
  50} no.~36, (2017) 364001}, \href{http://arxiv.org/abs/1703.10656}{{\ttfamily
  arXiv:1703.10656 [hep-th]}}.

\bibitem{Huang:2019fog}
K.-W. Huang, ``{Stress-tensor commutators in conformal field theories near the
  lightcone},'' \href{http://dx.doi.org/10.1103/PhysRevD.100.061701}{{\em Phys.
  Rev. D} {\bfseries 100} no.~6, (2019) 061701},
  \href{http://arxiv.org/abs/1907.00599}{{\ttfamily arXiv:1907.00599
  [hep-th]}}.

\bibitem{Kravchuk:2018htv}
P.~Kravchuk and D.~Simmons-Duffin, ``{Light-ray operators in conformal field
  theory},'' \href{http://dx.doi.org/10.1007/JHEP11(2018)102}{{\em JHEP}
  {\bfseries 11} (2018) 102}, \href{http://arxiv.org/abs/1805.00098}{{\ttfamily
  arXiv:1805.00098 [hep-th]}}.

\bibitem{Cordova:2018ygx}
C.~C\'ordova and S.-H. Shao, ``{Light-ray Operators and the BMS Algebra},''
  \href{http://dx.doi.org/10.1103/PhysRevD.98.125015}{{\em Phys. Rev. D}
  {\bfseries 98} no.~12, (2018) 125015},
  \href{http://arxiv.org/abs/1810.05706}{{\ttfamily arXiv:1810.05706
  [hep-th]}}.

\bibitem{Belin:2020lsr}
A.~Belin, D.~M. Hofman, G.~Mathys, and M.~T. Walters, ``{On the Stress Tensor
  Light-ray Operator Algebra},''
  \href{http://arxiv.org/abs/2011.13862}{{\ttfamily arXiv:2011.13862
  [hep-th]}}.

\bibitem{Gonzo:2020xza}
R.~Gonzo and A.~Pokraka, ``{Light-ray operators, detectors and gravitational
  event shapes},'' \href{http://dx.doi.org/10.1007/JHEP05(2021)015}{{\em JHEP}
  {\bfseries 05} (2021) 015}, \href{http://arxiv.org/abs/2012.01406}{{\ttfamily
  arXiv:2012.01406 [hep-th]}}.

\bibitem{Besken:2020snx}
M.~Be\c{s}ken, J.~De~Boer, and G.~Mathys, ``{On local and integrated
  stress-tensor commutators},''
  \href{http://dx.doi.org/10.1007/JHEP07(2021)148}{{\em JHEP} {\bfseries 21}
  (2021) 148}, \href{http://arxiv.org/abs/2012.15724}{{\ttfamily
  arXiv:2012.15724 [hep-th]}}.

\bibitem{Korchemsky:2021htm}
G.~P. Korchemsky and A.~Zhiboedov, ``{On the light-ray algebra in conformal
  field theories},'' \href{http://dx.doi.org/10.1007/JHEP02(2022)140}{{\em
  JHEP} {\bfseries 02} (2022) 140},
  \href{http://arxiv.org/abs/2109.13269}{{\ttfamily arXiv:2109.13269
  [hep-th]}}.

\bibitem{Huang:2020ycs}
K.-W. Huang, ``{Lightcone Commutator and Stress-Tensor Exchange in $d>2$
  CFTs},'' \href{http://dx.doi.org/10.1103/PhysRevD.102.021701}{{\em Phys. Rev.
  D} {\bfseries 102} no.~2, (2020) 021701},
  \href{http://arxiv.org/abs/2002.00110}{{\ttfamily arXiv:2002.00110
  [hep-th]}}.

\bibitem{Huang:2022vcs}
K.-W. Huang, ``{Approximate symmetries in d = 4 CFTs with an Einstein gravity
  dual},'' \href{http://dx.doi.org/10.1007/JHEP09(2022)053}{{\em JHEP}
  {\bfseries 09} (2022) 053}, \href{http://arxiv.org/abs/2202.09998}{{\ttfamily
  arXiv:2202.09998 [hep-th]}}.

\bibitem{Beem:2013sza}
C.~Beem, M.~Lemos, P.~Liendo, W.~Peelaers, L.~Rastelli, and B.~C. van Rees,
  ``{Infinite Chiral Symmetry in Four Dimensions},''
  \href{http://dx.doi.org/10.1007/s00220-014-2272-x}{{\em Commun. Math. Phys.}
  {\bfseries 336} no.~3, (2015) 1359--1433},
  \href{http://arxiv.org/abs/1312.5344}{{\ttfamily arXiv:1312.5344 [hep-th]}}.

\bibitem{Beem:2014kka}
C.~Beem, L.~Rastelli, and B.~C. van Rees, ``{$ \mathcal{W} $ symmetry in six
  dimensions},'' \href{http://dx.doi.org/10.1007/JHEP05(2015)017}{{\em JHEP}
  {\bfseries 05} (2015) 017}, \href{http://arxiv.org/abs/1404.1079}{{\ttfamily
  arXiv:1404.1079 [hep-th]}}.

\bibitem{Beem:2014rza}
C.~Beem, W.~Peelaers, L.~Rastelli, and B.~C. van Rees, ``{Chiral algebras of
  class S},'' \href{http://dx.doi.org/10.1007/JHEP05(2015)020}{{\em JHEP}
  {\bfseries 05} (2015) 020}, \href{http://arxiv.org/abs/1408.6522}{{\ttfamily
  arXiv:1408.6522 [hep-th]}}.

\bibitem{Johansen:1994ud}
A.~Johansen, ``{Infinite conformal algebras in supersymmetric theories on four
  manifolds},'' \href{http://dx.doi.org/10.1016/0550-3213(94)00408-7}{{\em
  Nucl. Phys. B} {\bfseries 436} (1995) 291--341},
  \href{http://arxiv.org/abs/hep-th/9407109}{{\ttfamily arXiv:hep-th/9407109}}.

\bibitem{Kapustin:2006hi}
A.~Kapustin, ``{Holomorphic reduction of N=2 gauge theories, Wilson-'t Hooft
  operators, and S-duality},''
  \href{http://arxiv.org/abs/hep-th/0612119}{{\ttfamily arXiv:hep-th/0612119}}.

\bibitem{Belavin:2017lvd}
V.~Belavin, Y.~Haraoka, and R.~Santachiara, ``{Rigid Fuchsian systems in
  2-dimensional conformal field theories},''
  \href{http://dx.doi.org/10.1007/s00220-018-3274-x}{{\em Commun. Math. Phys.}
  {\bfseries 365} no.~1, (2019) 17--60},
  \href{http://arxiv.org/abs/1711.04361}{{\ttfamily arXiv:1711.04361
  [hep-th]}}.

\bibitem{Ribault:2014hia}
S.~Ribault, ``{Conformal field theory on the plane},''
  \href{http://arxiv.org/abs/1406.4290}{{\ttfamily arXiv:1406.4290 [hep-th]}}.

\end{thebibliography}\endgroup

\end{document}